\documentclass[12pt]{iopart}

\usepackage{bm}
\usepackage{amssymb}
\usepackage{graphicx}
\usepackage{multirow}
\usepackage{color}
\usepackage{cite}

\usepackage{enumerate}

\usepackage[linktoc=all,unicode=true,pdfusetitle,  bookmarks=true,bookmarksnumbered=false,bookmarksopen=false, breaklinks=true,pdfborder={0 0 0},backref=false,colorlinks=true]{hyperref}
\hypersetup{linkcolor=blue,citecolor=red}

\usepackage[hyphenbreaks]{breakurl}

\newcommand{\binom}[2]{{{#1}\choose{#2}}}

\newcommand{\underset}[2]{\mathop{#2}\limits_{#1}}

\newcommand{\beq}{\begin{equation}}
\newcommand{\eeq}{\end{equation}}
\newcommand{\beqa}{\begin{eqnarray}}
\newcommand{\eeqa}{\end{eqnarray}}

\newcommand{\ex}{{\text{ex}}}
\newcommand{\id}{\text{id}}

\newcommand{\text}[1]{\mathrm{#1}}

\def\bal#1\eal{\begin{align}#1\end{align}}

\newcommand{\ed}[1]{\end{document}}

\newcommand{\calo}{{\cal O}}
\newcommand{\bq}{\begin{eqnarray}}
\newcommand{\eq}{\end{eqnarray}}

\eqnobysec

\begin{document}

\title[One-dimensional Janus fluids]{Finite-size effects and thermodynamic limit in one-dimensional Janus fluids}
\author{R Fantoni$^1$, M A G Maestre$^2$ and A Santos$^{2,3}$}
\
\address{$^1$Dipartimento di Fisica, Universit\`a di Trieste, Strada Costiera 11, 34151 Grignano (Trieste), Italy\\
$^2$Departamento de F\'isica, Universidad de Extremadura, E--06006 Badajoz, Spain\\
$^3$Instituto de Computaci\'on Cient\'ifica Avanzada (ICCAEx), Universidad de Extremadura, E--06006 Badajoz, Spain}

\eads{\mailto{riccardo.fantoni@posta.istruzione.it}, \mailto{maestre@unex.es}, \mailto{andres@unex.es}}

\date{\today}
\begin{abstract}
The equilibrium properties of a Janus fluid made of two-face particles confined to a one-dimensional channel are revisited. The exact Gibbs free energy for a finite number of particles $N$ is exactly derived for both quenched and annealed realizations. It is proved that the results for both classes of systems  tend in the thermodynamic limit ($N\to\infty$) to a common expression recently derived (Maestre M A G and Santos A 2020 J Stat Mech 063217). The theoretical finite-size results are particularized to the Kern--Frenkel model and confirmed by Monte Carlo simulations for quenched and (both biased and unbiased) annealed systems.
\end{abstract}
\noindent{\it Keywords\/}: exact results, classical Monte Carlo simulations,
colloids, bio-colloids and nano-colloids

 \maketitle

\section{Introduction\label{sec1}}

New materials chemical technology allows for the synthesis of colloidal-size particles with patches exhibiting an interaction pattern different from that of the rest of the surface \cite{RML05,WLZL08,WM13}. When the patch occupies a hemisphere, we are in the presence of so-called Janus particles \cite{WM13,BF01,SGP09,YHHD10,F13,OTSM15}.

\begin{figure}
\begin{center}
\includegraphics[width=.6\columnwidth]{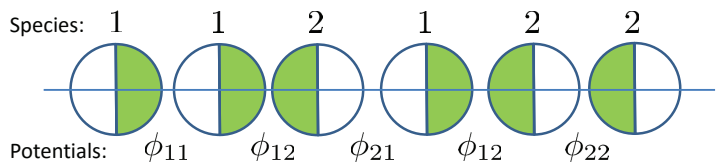}
\caption{Sketch of a binary mixture of one-dimensional Janus particles. Particles of species $1$ ($2$) have a white (green) left face and a green (white) right face. In general, four types of interactions are possible: green--white ($\phi_{11}$),  green--green ($\phi_{12}$),  white--white ($\phi_{21}$), and and white--green ($\phi_{22}$). 
However, in most of this paper we will
assume $\phi_{11} = \phi_{22} =\phi_{21}$.
In this particular example, $x_1=x_2=\frac{1}{2}$ and $N=6$.}
\label{sketch}
\end{center}
\end{figure}

One-dimensional fluids play an important role in statistical mechanics because they often offer integrable systems
\cite{HGM34,T36,N40a,N40b,T42,vH50,SZK53,K55b,LPZ62,KT68,LZ71,P76,P82,BOP87,K91,R91,HC04,S07,SFG08,BNS09,FGMS10,F10b,S14,S16,FS17,MS19}.
In a recent paper \cite{MS20}, two of us derived the exact equilibrium thermodynamic and  structural properties of one-dimensional Janus fluids in the thermodynamic limit (TL). The system consisted in a  binary mixture of two-face $N_i=x_i N$ particles of species $i=1,2$, where $x_i$ is the mole fraction of species $i$ and $N$ is the total number of particles. See figure \ref{sketch} for a sketch of the system.
In this type of systems (henceforth referred to as \emph{quenched}), the number of particles ($N_1$ and $N_2$) with each face orientation is kept fixed but of course one needs to average over all possible microscopic configurations to obtain macroscopic quantities.
Interestingly, the theoretical predictions for quenched systems agreed excellently well with Monte Carlo (MC) simulations for \emph{annealed} systems (where at each MC attempt a particle is assigned the face orientation $1$ or $2$ with probabilities $q_1$ and $q_2=1-q_1$, respectively) with $N=500$.

The investigation of reference \cite{MS20} stimulates a few questions: (i) can the exact derivation of the Gibbs free energy in the TL ($N\to\infty$) be extended to quenched and/or annealed finite-$N$ systems?; (ii) does the quenched$\leftrightarrow$annealed equivalence break down at finite $N$?; (iii) can those theoretical predictions be validated by MC simulations?;  (iv) is the dependence of the average mole fraction $\langle x_1\rangle$ on the probability $q_1$  robust with respect to $N$ in annealed MC simulations for biased situations ($q_1\neq \frac{1}{2}$)? The main aim of this paper is to address those questions. As will be seen, the answers are affirmative in all the cases.

The remainder of this paper is organized as follows. Section \ref{sec2} presents the derivation of the configuration integral, and hence of the Gibbs free energy $G$, for a finite-size quenched binary mixture in the isothermal-isobaric ensemble. Those results are then used in section \ref{sec3} to derive $G$ for an annealed fluid. Since the exact results in sections \ref{sec2} and \ref{sec3} apply to any choice of the two nearest-neighbor interaction potentials $\phi_{11}=\phi_{22}=\phi_{21}$ and $\phi_{12}$ (see figure \ref{sketch}), the expressions are particularized in section \ref{sec4} to the Kern--Frenkel model \cite{KF03}, where $\phi_{11}$ and $\phi_{12}$ are the hard-rod and square-well potentials, respectively. The theoretical results are validated and confirmed by MC simulations in section \ref{sec5}, where also the case of biased annealed systems is addressed. Finally, the main results of the work are summarized in \ref{sec6}. The most technical parts of the paper are relegated to five appendices.

\section{Finite-$N$ Gibbs free energy of a quenched binary mixture of Janus rods}
\label{sec2}
\subsection{The system}
Let us consider a one-dimensional binary fluid mixture made of $N_1$ particles of species $1$ (right `spin') and $N_2=N-N_1$ particles  of species $2$ (left `spin')  on a line of length $L$ (see figure \ref{sketch}).
Henceforth, we will use Latin and Greek indices for species and particles, respectively. A particular spatial configuration will be denoted as $\bm{x}\equiv \{x_\alpha;\alpha=1,2,\ldots,N\}$. Analogously, a particular spin (or species) configuration will be denoted as $\bm{s}\equiv \{s_\alpha;\alpha=1,2,\ldots,N\}$, where $s_\alpha=1,2$ represents the spin of particle $\alpha$. Since we are considering a quenched mixture, the number of possible spin configurations are restricted by the constraint
\begin{equation}
\label{constr}
\sum_{\alpha=1}^N \delta_{s_\alpha,1}=N_1.
\end{equation}
The total number of allowed spin configurations is $\binom{N}{N_1}$.

We assume that the rods are impenetrable and that their interaction is restricted to nearest neighbors. Given $\bm{s}$ and $\bm{x}$, the  total potential energy can be written as
\begin{equation}
\Phi_N(\bm{s},\bm{x})=\sum_{\alpha=1}^{N-1}\phi_{s_\alpha,s_{\alpha+1}}(x_{\alpha+1}-x_\alpha)+\omega \phi_{s_N,s_{1}}(x_1+L-x_N),
\label{N1}
\end{equation}
where, without loss of generality, we assume that particles $1,2,\ldots,N$ are ordered from left to right. In equation \eref{N1}, $\omega=1$ if periodic boundary conditions are applied and $\omega=0$ otherwise (open systems).

\begin{figure}
\begin{center}
\includegraphics[width=.8\columnwidth]{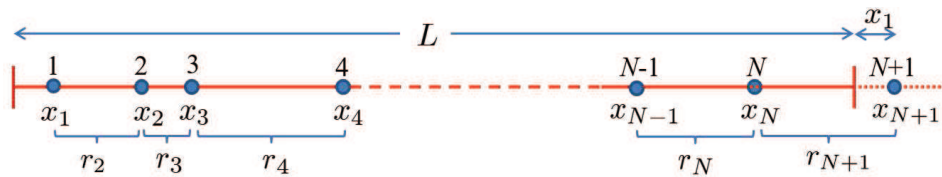}
\caption{Illustration of the change of variables \eref{ri}.}
\label{pbc}
\end{center}
\end{figure}

\subsection{Isothermal-isobaric partition function}
In the isothermal-isobaric ensemble, the  partition function is \cite{HM13,S16}
\begin{equation}
\label{Ztotal}
\mathcal{Z}_{N_1,N_2}(\beta,\gamma)=\mathcal{Z}_{N_1,N_2}^\id(\beta,\gamma)\mathcal{Q}_{N_1,N_2}(\beta,\gamma),
\end{equation}
where
\begin{equation}
\label{Z_id}
\fl
\mathcal{Z}_{N_1,N_2}^\id(\beta,\gamma)=\frac{C_{N_1,N_2}}{L_{\text{ref}}\left[\Lambda_1(\beta)\right]^{N_1}\left[\Lambda_2(\beta)\right]^{N_2}},\quad C_{N_1,N_2}(\gamma)\equiv\binom{N}{N_1}\gamma^{-(N+1)},
\end{equation}
is the ideal-gas partition function and
\begin{equation}
\label{QN}
\mathcal{Q}_{N_1,N_2}(\beta,\gamma)=\frac{1}{C_{N_1,N_2}(\gamma)}\sum_{\bm{s}}'\int_0^\infty\rmd L \, \rme^{-\gamma L}\underset{0<x_1<\cdots<x_N<L}{\int \rmd^N \bm{x}} \,\rme^{-\beta \Phi_N(\bm{s},\bm{x})}
\end{equation}
is the configuration integral.
Here, $\beta\equiv 1/k_{\text{B}}T$ ($k_{\text{B}}$ and $T$ being the Boltzmann constant and the absolute temperature, respectively) and $\gamma\equiv \beta p$ ($p$ being the pressure). In equation \eref{Z_id}, $L_{\text{ref}}$ is a reference length (introduced to make $\mathcal{Z}_{N}^\id$ dimensionless) and $\Lambda_i(\beta)\equiv h\sqrt{{\beta}/{2\pi m_i}}$ is the thermal de Broglie wavelength ($h$ being the Planck constant and $m_i$ being the mass of a particle of species $i$). In equation \eref{QN}, the prime in the summation denotes the constraint \eref{constr}. Note that, by construction, $\mathcal{Q}_{N_1,N_2}=1$ if $\Phi_N=0$.

Let us make $\mathcal{Q}_{N_1,N_2}$ more explicit. First,
\begin{eqnarray}
\label{QN_a}
\fl
\mathcal{Q}_{N_1,N_2}&=&\frac{1}{C_{N_1,N_2}}\sum_{\bm{s}}'\int_0^\infty\rmd L \, \rme^{-\gamma L}\int_0^L\rmd x_1\int_{x_1}^L\rmd x_2\cdots \int_{x_{N-1}}^L\rmd x_N\, \rme^{-\beta \Phi_N(\bm{s},\bm{x})}\nonumber\\
\fl
&=&\frac{1}{C_{N_1,N_2}}\sum_{\bm{s}}'\int_0^\infty\rmd x_1\int_{x_1}^\infty\rmd x_2\cdots \int_{x_{N-1}}^\infty\rmd x_N\int_{x_N}^\infty\rmd L \, \rme^{-\gamma L-\beta \Phi_N(\bm{s},\bm{x})},
\end{eqnarray}
where in the second step we have changed the order of integration. Next, we perform the change of variables $\{x_1,x_2,\ldots,x_N,L\}\to\{x_1,r_2,\ldots,r_N,r_{N+1}\}$, where (see figure \ref{pbc})
\begin{equation}
\label{ri}
r_i\equiv x_{i}-x_{i-1} \quad (i=2,\ldots,N),\quad r_{N+1}\equiv x_1+L-x_N.
\end{equation}
Note that $L=\sum_{\alpha=2}^{N+1}r_\alpha$. With this change of variables, equation \eref{QN_a} becomes
\begin{eqnarray}
\label{QN_b}
\mathcal{Q}_{N_1,N_2}&=&\frac{1}{C_{N_1,N_2}}\sum_{\bm{s}}'\left[
\prod_{\alpha=2}^N\int_0^\infty\rmd r_\alpha \, \rme^{-\gamma r_\alpha-\beta\phi_{s_{\alpha-1},s_\alpha}(r_\alpha)}
\right]\nonumber\\
&&\times
\int_0^\infty\rmd x_1\int_{x_1}^\infty\rmd r_{N+1}\rme^{-\gamma r_{N+1}-\beta\omega\phi_{s_{N},s_1}(r_{N+1})}\nonumber\\
&=&\frac{1}{C_{N_1,N_2}}\sum_{\bm{s}}'\left[\prod_{\alpha=2}^N\Omega_{s_{\alpha-1},s_\alpha}(\beta,\gamma)\right]\left[-\frac{\partial \Omega_{s_{N},s_1}(\beta\omega,\gamma)}{\partial\gamma}\right],
\end{eqnarray}
where
\begin{equation}
\Omega_{ij}(\beta,\gamma)\equiv\int_0^\infty \rmd r\, \rme^{-\gamma r-\beta\phi_{ij}(r)}.
\label{35.1}
\end{equation}

Henceforth, we particularize to open systems ($\omega=0$), so that
\begin{equation}
\label{QN_c}
\mathcal{Q}_{N_1,N_2}=\frac{\gamma^{-2}}{C_{N_1,N_2}}\sum_{\bm{s}}'\prod_{\alpha=2}^N\Omega_{s_{\alpha-1},s_\alpha}.
\end{equation}
Given a spin configuration $\bm{s}$, let us call $n_{ij}(\bm{s})$ the number of pairs $ij$. Thus,
\begin{equation}
\prod_{\alpha=2}^N\Omega_{s_{\alpha-1},s_\alpha}=\Omega_{11}^{n_{11}(\bm{s})}\Omega_{22}^{n_{22}(\bm{s})}\Omega_{12}^{n_{12}(\bm{s})}\Omega_{21}^{n_{21}(\bm{s})}.
\end{equation}
Obviously, $n_{11}+n_{22}+n_{12}+n_{21}=N-1$. If we call $w(n_{11},n_{22},n_{12},n_{21})$ the number of spin configurations with $n_{ij}$ pairs $ij$, equation \eref{QN_c} can be rewritten as
\begin{equation}
\label{QN_d}
\mathcal{Q}_{N_1,N_2}=\frac{\gamma^{-2}}{C_{N_1,N_2}}\sum_{n_{11},n_{22},n_{12},n_{21}}w(n_{11},n_{22},n_{12},n_{21})\Omega_{11}^{n_{11}}\Omega_{22}^{n_{22}}\Omega_{12}^{n_{12}}\Omega_{21}^{n_{21}}.
\end{equation}
Table \ref{table1} shows the possible values of $n_{ij}$ and $w$ for the simple example of $N_1=4$ and $N_2=2$.

\begin{table}
\caption{\label{table1}Spin configurations $\bm{s}$ for $N_1=4$ and $N_2=2$, organized according to the number ($n_{ij}$)  of pairs $ij$.
The number of spin configurations sharing the same values of $n_{ij}$ is given by $w(\{n_{ij}\})$; analogously, $w_{12}(n_{12})$ is the number of spin configurations sharing the same $n_{12}$, regardless of the values of $n_{11}$, $n_{22}$, and $n_{21}$.}
\begin{indented}
\item[]
\begin{tabular}{@{}ccccccc}
\br
$n_{11}$&$n_{22}$&$n_{12}$&$n_{21}$&$\bm{s}$&$w$&$w_{12}$\\
\mr
$3$&$1$&$0$&$1$&$\{221111\}$&$1$&$\}1$\\
$3$&$1$&$1$&$0$&$\{111122\}$&$1$&\multirow{4}{*}{$\left\}\rule{0cm}{1cm}\right.8$}\\
$3$&$0$&$1$&$1$&$\{211112\}$&$1$&\\
$2$&$1$&$1$&$1$&$\{111221\}, \{112211\}, \{122111\}$&$3$&\\
$2$&$0$&$1$&$2$&$\{211121\},\{211211\}, \{212111\}$&$3$&\\
$2$&$0$&$2$&$1$&$\{111212\},\{112112\}, \{121112\}$&$3$&\multirow{2}{*}{$\left\}\rule{0cm}{0.5cm}\right.6$}\\
$1$&$0$&$2$&$2$&$\{112121\},\{121121\}, \{121211\}$&$3$&\\
\br
\end{tabular}
\end{indented}
\end{table}

In general, the evaluation of the number of combinations $w(\{n_{ij}\})$ is quite hard. On the other hand, since in the end we will apply the results to the Kern--Frenkel Janus model \cite{KF03},
we can particularize to the case where $\phi_{11}(r)=\phi_{22}(r)=\phi_{21}(r)$, what implies $\Omega_{11}=\Omega_{22}=\Omega_{21}$, so that equation \eref{QN_d} reduces to
\begin{equation}
\label{QN_e}
\mathcal{Q}_{N_1,N_2}=\frac{\gamma^{N-1}}{\binom{N}{N_1}}\sum_{n_{12}=0}^{\min\{N_1,N_2\}}w_{12}(n_{12})\Omega_{11}^{N-1-n_{12}}\Omega_{12}^{n_{12}},
\end{equation}
where  $w_{12}(n_{12})$ stands for the number of spin configurations with $n_{12}$ pairs $12$.

To determine $w_{12}(n_{12})$, imagine that we enumerate particles of each species $i=1$ and $2$ from left to right as $\alpha_i=1,\ldots,N_i$. Then, each pair of type $12$ can be identified with a label $(\alpha_1,\alpha_2)$. Thus, given a number $n_{12}$, each compatible spin configuration $\bm{s}$ is characterized  by $n_{12}$ pairs of the form $(\alpha_1,\alpha_2)$. For example, if $N_1=4$ and $N_2=2$ (table \ref{table1}), the spin configuration $\bm{s}=\{112121\}$ has $n_{12}=2$ pairs: $(\alpha_1,\alpha_2)=(2,1)$ and $(3,2)$, while the
spin configuration $\bm{s}=\{211121\}$ has a single $n_{12}$ pair: $(\alpha_1,\alpha_2)=(3,2)$. There is a one-to-one correspondence between the $n_{12}$ pairs of the form $(\alpha_1,\alpha_2)$ and the associated spin configuration $\bm{s}$. As a consequence, the number of spin configurations $w_{12}(n_{12})$ with $n_{12}$ pairs of type $12$ is given by the number of ways of choosing  the $n_{12}$ labels $\alpha_1$ out of $N_1$  possible values and the $n_{12}$ labels $\alpha_2$ out of $N_2$ possible values.
Therefore,
\begin{equation}
w_{12}(n_{12})=\binom{N_1}{n_{12}}\binom{N_2}{n_{12}}.
\end{equation}
As a test of consistency, note that the total number of spin configurations is recovered as $\sum_{n_{12}=0}^{\min\{N_1,N_2\}}w_{12}(n_{12})=\binom{N}{N_1}$.
Finally, the configuration integral is
\begin{equation}
\label{QN_f}
\mathcal{Q}_{N_1,N_2}=\frac{\left(\gamma \Omega_{11}\right)^{N-1}}{\binom{N}{N_1}}\Xi_{N_1,N_2},\quad \Xi_{N_1,N_2}\equiv \sum_{n=0}^{\min\{N_1,N_2\}}\xi_{N_1,N_2}(n),
\end{equation}
where
\begin{equation}
\label{xi}
\xi_{N_1,N_2}(n)\equiv\binom{N_1}{n}\binom{N_2}{n}\left(1-R\right)^{-n},\quad R\equiv 1-\frac{\Omega_{11}}{\Omega_{12}}.
\end{equation}
Interestingly, $\Xi_{N_1,N_2}$ can be formally rewritten in terms of the hypergeometric function:
\begin{equation}
\label{hyper}
\Xi_{N_1,N_2}=_2\!\!F_1\left(-N_1,-N_2;1,\frac{1}{1-R}\right).
\end{equation}

\subsection{Gibbs free energy, internal energy, and equation of state}
The finite-size  Gibbs free energy $G_N(T,p,x_1)$ is related to the partition function $\mathcal{Z}_{N_1,N_2}(\beta,\gamma)$ as $G_N=-k_\text{B}T\ln \mathcal{Z}_{N_1,N_2}$ \cite{HM13,S16}. According to equations \eref{Ztotal}, \eref{Z_id}, and \eref{QN_f}, the finite-size Gibbs free energy per particle $g_N=G_N/N$ can be decomposed as $g_N=g_N^\id+g_N^\ex$, with
\numparts
\begin{eqnarray}
\label{gidN}
\beta g_N^\id&=&x_1\ln\left(\gamma \Lambda_1\right)+x_2\ln\left(\gamma \Lambda_2\right)
-N^{-1}\ln\binom{N}{N_1}+N^{-1}\ln\left(\gamma L_{\text{ref}}\right),\\
\label{gexN}
\beta g_N^\ex&=&-\left(1-N^{-1}\right)\ln\left(\gamma \Omega_{11}\right)-N^{-1}\ln\frac{\Xi_{N_1,N_2}}{\binom{N}{N_1}}.
\end{eqnarray}
\endnumparts

By viewing $g_N$ as a function of $\beta$ and $\gamma$ (instead of as a function of $T$ and $p$), it is easy to obtain the average volume (length) per particle ($v_N$) and the excess energy per particle ($u_N$) at finite $N$ as
\begin{equation}
\label{6}
v_N=\left(\frac{\partial \beta g_N}{\partial \gamma}\right)_{\beta }=v_N^\id+v_N^\ex,\quad
u_N=\left(\frac{\partial \beta g_N}{\partial \beta}\right)_{\gamma}=u^\id+u_N^\ex.
\end{equation}
From equations \eref{gidN} and \eref{gexN}, one has
\numparts
\begin{eqnarray}
\label{vidN}
v_N^\id&=&\frac{1+N^{-1}}{\gamma},\quad
u^\id=\frac{1}{2\beta},\\
\label{vexN}
v_N^\ex&=&-\left(1-N^{-1}\right)\left(\frac{\partial\ln\left(\gamma \Omega_{11}\right)}{\partial\gamma}\right)_\beta-N^{-1}\frac{\partial\ln\Xi_{N_1,N_2}}{\partial R}\left(\frac{\partial R}{\partial\gamma}\right)_\beta,\\
\label{uexN}
u_N^\ex&=&-\left(1-N^{-1}\right)\left(\frac{\partial\ln\Omega_{11}}{\partial\beta}\right)_\gamma-N^{-1}\frac{\partial\ln\Xi_{N_1,N_2}}{\partial R}\left(\frac{\partial R}{\partial\beta}\right)_\gamma,
\end{eqnarray}
\endnumparts
where, in view of equation \eref{hyper},
\begin{equation}
\label{hyper2}
\frac{\partial\Xi_{N_1,N_2}}{\partial R}=\frac{N_1N_2}{(1-R)^2}\, _2F_1\left(-N_1+1,-N_2+1;2,\frac{1}{1-R}\right).
\end{equation}

\subsection{Limit $N\to\infty$}

Equations \eref{gexN}, \eref{vexN}, and \eref{uexN} provide the excess quantities for any finite $N$. It is important to take the limit $N\to\infty$ to obtain the TL expressions and their first finite-$N$ corrections.

In \ref{appA} it is proved that, for large $N$ at fixed mole fractions,
\begin{equation}
\label{Xi_N1N2_c}
\Xi_{N_1,N_2}\approx \frac{\rme^{N\bar{\psi}_0}}{\sqrt{2\pi N y_0(2-y_0/x_1x_2)}},
\end{equation}
where
\begin{equation}
\label{psi00}
\bar{\psi}_0=-x_1\ln\left(1-\frac{y_0}{x_1}\right)-x_2\ln\left(1-\frac{y_0}{x_2}\right),\quad y_0= \frac{1-\sqrt{1-4x_1 x_2R}}{2R}.
\end{equation}
As a consistency test, note that in the case of equal interactions ($R\to 0$), one has $y_0\to x_1 x_2$ and $\bar{\psi}_0\to -x_1\ln x_1-x_2\ln x_2$, so that $\Xi_{N_1,N_2}\to (x_1^{N_1}x_2^{N_2}\sqrt{2\pi N x_1 x_2})^{-1}$. The latter expression is not but the Stirling approximation of $\binom{N}{N_1}$, as it should be.

Thus, from equation \eref{gexN} we obtain
\begin{equation}
\label{gNapprox_a}
\beta g_N^\ex\approx \beta g_{\text{TL}}^\ex+N^{-1}\ln\left[\gamma \Omega_{11}\sqrt{(2-y_0/x_1x_2)y_0/x_1x_2}\right],
\end{equation}
where
\begin{equation}
\label{gTLa}
\beta g_{\text{TL}}^\ex=-\ln(\gamma\Omega_{11})-\bar{\psi}_0-x_1\ln x_1-x_2\ln x_2
\end{equation}
and we have taken into account that $N^{-1}\ln\binom{N}{N_1}\approx -x_1\ln x_1-x_2\ln x_2-N^{-1}\ln\sqrt{2\pi N x_1 x_2}$.
Obviously, $g_{\text{TL}}^\ex$ is the excess Gibbs free energy per particle in the TL. That quantity was evaluated by a completely independent route in reference \cite{MS20} with the result
\begin{equation}
\label{gTLb}
\fl
\beta g_{\text{TL}}^\ex=-\ln(\gamma\Omega_{11})
-\ln\frac{1+\sqrt{1-4x_1x_2R}}{2\sqrt{1-R}}
+|x_1-x_2|\ln\frac{|x_1-x_2|+\sqrt{1-4x_1x_2R}}{(|x_1-x_2|+1)\sqrt{1-R}}.
\end{equation}
Taking into account the identity (see \ref{appB} for a proof)
\begin{eqnarray}
\label{proof}
\bar{\psi}_0&=&-x_1 \ln x_1-x_2\ln x_2
+\ln\frac{1+\sqrt{1-4x_1x_2R}}{2\sqrt{1-R}}
\nonumber\\&&
-|x_1-x_2|\ln\frac{|x_1-x_2|+\sqrt{1-4x_1x_2R}}{(|x_1-x_2|+1)\sqrt{1-R}},
\end{eqnarray}
it is obvious that equations \eref{gTLa} and \eref{gTLb} are equivalent. Note, however, that equation \eref{gTLa} is more compact than equation \eref{gTLb}.

As for the average volume and internal energy per particle, application of equation \eref{6} yields
\numparts
\begin{eqnarray}
\label{vTL}
v_{\text{TL}}^\ex&=&-\left(\frac{\partial\ln\left(\gamma \Omega_{11}\right)}{\partial\gamma}\right)_\beta-\frac{y_0^3/x_1 x_2}{(1-y_0/x_1)(1-y_0/x_2)}\left(\frac{\partial R}{\partial \gamma}\right)_{\beta},\\
\label{uTL}
u_{\text{TL}}^\ex&=&-\left(\frac{\partial\ln \Omega_{11}}{\partial\beta}\right)_\gamma-\frac{y_0^3/x_1 x_2}{(1-y_0/x_1)(1-y_0/x_2)}\left(\frac{\partial R}{\partial \beta}\right)_{\gamma},
\end{eqnarray}
\begin{eqnarray}
\label{4}
\fl
v_N^\ex- v_{\text{TL}}^\ex&\approx& N^{-1}\left(\frac{\partial\ln\left(\gamma \Omega_{11}\right)}{\partial\gamma}\right)_\beta+\frac{N^{-1}}{2}\frac{(1-y_0/x_1 x_2)y_0^2/2x_1 x_2}{(1-y_0/2x_1x_2)^2}\left(\frac{\partial R}{\partial \gamma}\right)_{\beta},
\\
\label{5}
\fl
u_N^\ex-u_{\text{TL}}^\ex&\approx &N^{-1}\left(\frac{\partial\ln\Omega_{11}}{\partial\beta}\right)_\gamma+\frac{N^{-1}}{2}\frac{(1-y_0/x_1 x_2)y_0^2/2x_1 x_2}{(1-y_0/2x_1x_2)^2}\left(\frac{\partial R}{\partial \beta}\right)_{\gamma}.
\end{eqnarray}
\endnumparts
Note that, while $u^\id$ has no finite-$N$ contribution, this is not so for $v_N^\id$. According to equation \eref{vidN}, $v_N^\id=v_\text{TL}^\id+\left(\gamma N\right)^{-1}$, with $v_\text{TL}^\id=\gamma^{-1}$.

\subsection{Equimolar mixture}
In the special case of an equimolar binary mixture ($x_1=x_2=\frac{1}{2}$), equations \eref{gTLa}, \eref{vTL}, and \eref{uTL} become
\numparts
\begin{eqnarray}
\label{gTLequi}
\beta g_{\text{TL}}^\ex&=&-\ln\left[\frac{\gamma \Omega_{11}}{2}\left(1+\frac{1}{\sqrt{1-R}}\right)\right],\\
\label{vTLequi}
v_{\text{TL}}^\ex&=&-\left(\frac{\partial\ln\left(\gamma \Omega_{11}\right)}{\partial\gamma}\right)_\beta-\frac{1-\sqrt{1-R}}{2R(1-R)}\left(\frac{\partial R}{\partial \gamma}\right)_{\beta},\\
\label{uTLequi}
u_{\text{TL}}^\ex&=&-\left(\frac{\partial\ln \Omega_{11}}{\partial\beta}\right)_\gamma-\frac{1-\sqrt{1-R}}{2R(1-R)}\left(\frac{\partial R}{\partial \beta}\right)_{\gamma}.
\end{eqnarray}
\endnumparts
Analogously, equations \eref{gNapprox_a}, \eref{4}, and \eref{5} simplify to
\numparts
\begin{eqnarray}
\label{gNequi}
g_N^\ex-g_{\text{TL}}^\ex&\approx&N^{-1}\ln\left[2\gamma\Omega_{11}\frac{\left(1-\sqrt{1-R}\right)\left(1-R\right)^{1/4}}{R}\right],\\
\label{4b}
v_N^\ex- v_{\text{TL}}^\ex&\approx& N^{-1}\left(\frac{\partial\ln\left(\gamma \Omega_{11}\right)}{\partial\gamma}\right)_\beta-{N^{-1}}\frac{\left(1-\sqrt{1-R}\right)^2}{4R(1-R)}\left(\frac{\partial R}{\partial \gamma}\right)_{\beta},
\\
\label{5b}
u_N^\ex-u_{\text{TL}}^\ex&\approx &N^{-1}\left(\frac{\partial\ln \Omega_{11}}{\partial\beta}\right)_\gamma-{N^{-1}}\frac{\left(1-\sqrt{1-R}\right)^2}{4R(1-R)}\left(\frac{\partial R}{\partial \beta}\right)_{\gamma}.
\end{eqnarray}
\endnumparts

\section{Finite-$N$ Gibbs free energy of annealed Janus fluids}
\label{sec3}
In the case of (unbiased) annealed systems, the total number of particles ($N$) is fixed but the number of particles ($N_1$ or $N_2$) with either spin orientation species is allowed to take any value between $0$ and $N$. Thus, the associated configuration integral is
\begin{equation}
\mathcal{Q}_{N}(\beta,\gamma)=\frac{1}{C_{N}(\gamma)}\sum_{N_1=0}^N\sum_{\bm{s}}'\int_0^\infty\rmd L \, \rme^{-\gamma L}\underset{0<x_1<\cdots<x_N<L}{\int \rmd^N \bm{x}} \,\rme^{-\beta \Phi_N(\bm{s},\bm{x})},
\end{equation}
where now $C_N(\gamma)=\sum_{N_1=0}^NC_{N_1,N_2}=2^N\gamma^{-(N+1)}$ to guarantee that $\mathcal{Q}_{N}=1$ if $\Phi_N=0$.

By following the same steps as those followed to arrive to equation \eref{QN_f}, we now get
\begin{equation}
\label{QN_ann}
\mathcal{Q}_{N}=\frac{\left(\gamma \Omega_{11}\right)^{N-1}}{2^N}\Xi_{N}, \quad \Xi_N\equiv \sum_{N_1=0}^N\Xi_{N_1,N_2}.
\end{equation}
Consequently,
\numparts
\begin{eqnarray}
\beta g_N^\ex&=&-\left(1-N^{-1}\right)\ln\left(\gamma \Omega_{11}\right)+\ln 2-N^{-1}\ln\Xi_N,\\
\label{vexann}
v_N^\ex&=&-\left(1-N^{-1}\right)\left(\frac{\partial\ln\left(\gamma \Omega_{11}\right)}{\partial\gamma}\right)_\beta-N^{-1}\frac{\partial\ln\Xi_N}{\partial R}\left(\frac{\partial R}{\partial\gamma}\right)_\beta,\\
\label{uexann}
u_N^\ex&=&-\left(1-N^{-1}\right)\left(\frac{\partial\ln\Omega_{11}}{\partial\beta}\right)_\gamma-N^{-1}\frac{\partial\ln\Xi_N}{\partial R}\left(\frac{\partial R}{\partial\beta}\right)_\gamma,
\end{eqnarray}
\endnumparts
where we recall that the quantity $R$ is defined by the second equality in equation \eref{xi}.

In the limit of large $N$ it is proved in \ref{appC} that
\begin{equation}
\label{sumXi}
\Xi_N\approx \left(1+\frac{1}{\sqrt{1-R}}\right)^N\frac{1+\sqrt{1-R}}{2}.
\end{equation}
Therefore,
\numparts
\begin{eqnarray}
\label{gNequi_b}
\beta g_N^\ex-\beta g_{\text{TL}}^\ex&\approx& N^{-1}\ln\frac{2\gamma\Omega_{11}}{1+\sqrt{1-R}},\\
\label{4c}
v_N^\ex- v_{\text{TL}}^\ex&\approx& N^{-1}\left(\frac{\partial\ln\left(\gamma \Omega_{11}\right)}{\partial\gamma}\right)_\beta+{N^{-1}}\frac{1-\sqrt{1-R}}{2R\sqrt{1-R}}\left(\frac{\partial R}{\partial \gamma}\right)_{\beta},
\\
\label{5c}
u_N^\ex-u_{\text{TL}}^\ex&\approx &N^{-1}\left(\frac{\partial\ln \Omega_{11}}{\partial\beta}\right)_\gamma+{N^{-1}}\frac{1-\sqrt{1-R}}{2R\sqrt{1-R}}\left(\frac{\partial R}{\partial \beta}\right)_{\gamma},
\end{eqnarray}
\endnumparts
where the TL quantities are given by equations \eref{gTLequi}--\eref{uTLequi}.

Comparison between equations \eref{gNequi}--\eref{5b} and equations \eref{gNequi_b}--\eref{5c} shows that, although the quenched and annealed systems are equivalent in the TL, they differ in their respective finite-size corrections.

\section{Particularization to the Kern--Frenkel model}
\label{sec4}
Thus far,  except for the constraint to nearest neighbors, the interaction potentials $\phi_{11}(r)$ and $\phi_{12}(r)$ are arbitrary. In the special case of \emph{isotropic} interactions, one has $\phi_{11}(r)=\phi_{12}(r)$, so that $R=0$. In that case,
\numparts
\begin{eqnarray}
\fl
\Xi_{N_1,N_2}&=& \binom{N_1}{N_2},\quad \Xi_{N}=2^N,\quad \mathcal{Q}_{N_1,N_2}=\mathcal{Q}_N=\left(\gamma\Omega_{11}\right)^{N-1},\\
\fl
\beta g_N^\ex&=&-\left(1-N^{-1}\right)\ln\left(\gamma \Omega_{11}\right),\\
\fl
v_N^\ex&=&-\left(1-N^{-1}\right)\left(\frac{\partial\ln\left(\gamma \Omega_{11}\right)}{\partial \gamma}\right)_\beta,\quad
u_N^\ex=-\left(1-N^{-1}\right)\left(\frac{\partial\ln \Omega_{11}}{\partial \beta}\right)_\gamma.
\end{eqnarray}
\endnumparts
Thus, the finite-size effects become almost trivial if the interactions are isotropic and, of course, no distinction between quenched and annealed systems remains.

The situation becomes much more interesting in the genuine Janus case $\phi_{11}(r)\neq \phi_{12}(r)$. We take now the well-known Kern--Frenkel model
\cite{KF03,FGSP11,F12,F13,MFGS13,FGMS13}, in which case $\phi_{11}(r)$ and $\phi_{12}(r)$ correspond to the hard-rod and square-well potentials, respectively, i.e.,
\begin{equation}
\fl
\phi_{11}(r)=
\left\{
\begin{array}{ll}
\infty,& r<\sigma,\\0,&r>\sigma,
\end{array}
\right.
\qquad
\phi_{12}(r)=
\left\{
\begin{array}{ll}
 \infty,& r<\sigma,\\
-\epsilon,&\sigma<r<\lambda\sigma,\\
0,&r>\lambda\sigma,
\end{array}
\right.
\label{KF}
\end{equation}
where $\lambda\leq 2$.
Henceforth, we take $\sigma=1$, $\epsilon=1$, and $\epsilon/k_{\text{B}}=1$ as units of length, energy, and temperature, respectively.
Therefore,
\numparts
\begin{equation}
\fl
\Omega_{11}=\frac{\rme^{-\gamma}}{\gamma},\quad \Omega_{12}=\rme^\beta\frac{\rme^{-\gamma}}{\gamma}-\left(\rme^\beta-1\right)\frac{\rme^{-\lambda\gamma}}{\gamma},
\quad R=\left\{1+\frac{1}{\left(\rme^{\beta}-1\right)\left[1-\rme^{-(\lambda-1)\gamma}\right]}\right\}^{-1},
\label{44}
\end{equation}
\begin{equation}
\left(\frac{\partial\ln\left(\gamma \Omega_{11}\right)}{\partial \gamma}\right)_\beta=-1,\quad
\left(\frac{\partial\ln \Omega_{11}}{\partial \beta}\right)_\gamma=0,
\end{equation}
\begin{equation}
\fl
\left(\frac{\partial R}{\partial\gamma}\right)_\beta=\left(1-R\right)^2\left(\rme^{\beta}-1\right)(\lambda-1)\rme^{-(\lambda-1)\gamma},\quad
\left(\frac{\partial R}{\partial\beta}\right)_\gamma=\left(1-R\right)^2\rme^{\beta}\left[1-\rme^{-(\lambda-1)\gamma}\right].
\end{equation}
\endnumparts

\section{Monte Carlo simulations}
\label{sec5}

\subsection{Equimolar quenched and unbiased annealed systems}

In order to confirm the theoretical results provided by equations \eref{vexN} and \eref{uexN} for quenched systems and by equations \eref{vexann} and \eref{uexann} for (unbiased) annealed systems, we have performed isothermal-isobaric Monte Carlo (MC) simulations. To make contact between the annealed and quenched results in the TL, we have considered equimolar mixtures ($x_1=\frac{1}{2}$) in the latter case.
Moreover, the Kern--Frenkel model \eref{KF} with $\lambda=1.2$ is chosen.
Some technical details about the simulation method are given in a \ref{appMC}.

\begin{figure}
\begin{center}
\includegraphics[width=.5\columnwidth]{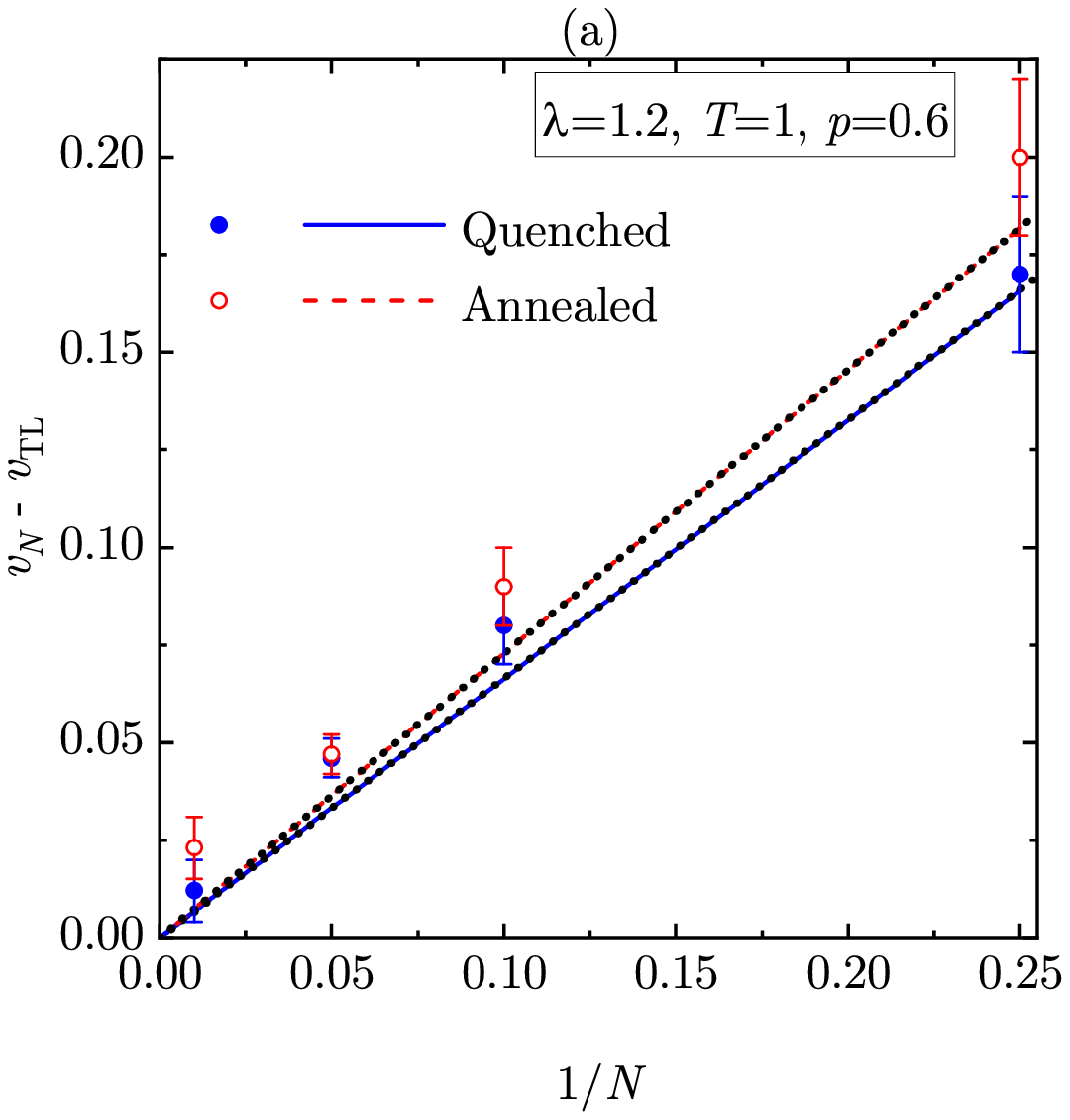}\includegraphics[width=.5\columnwidth]{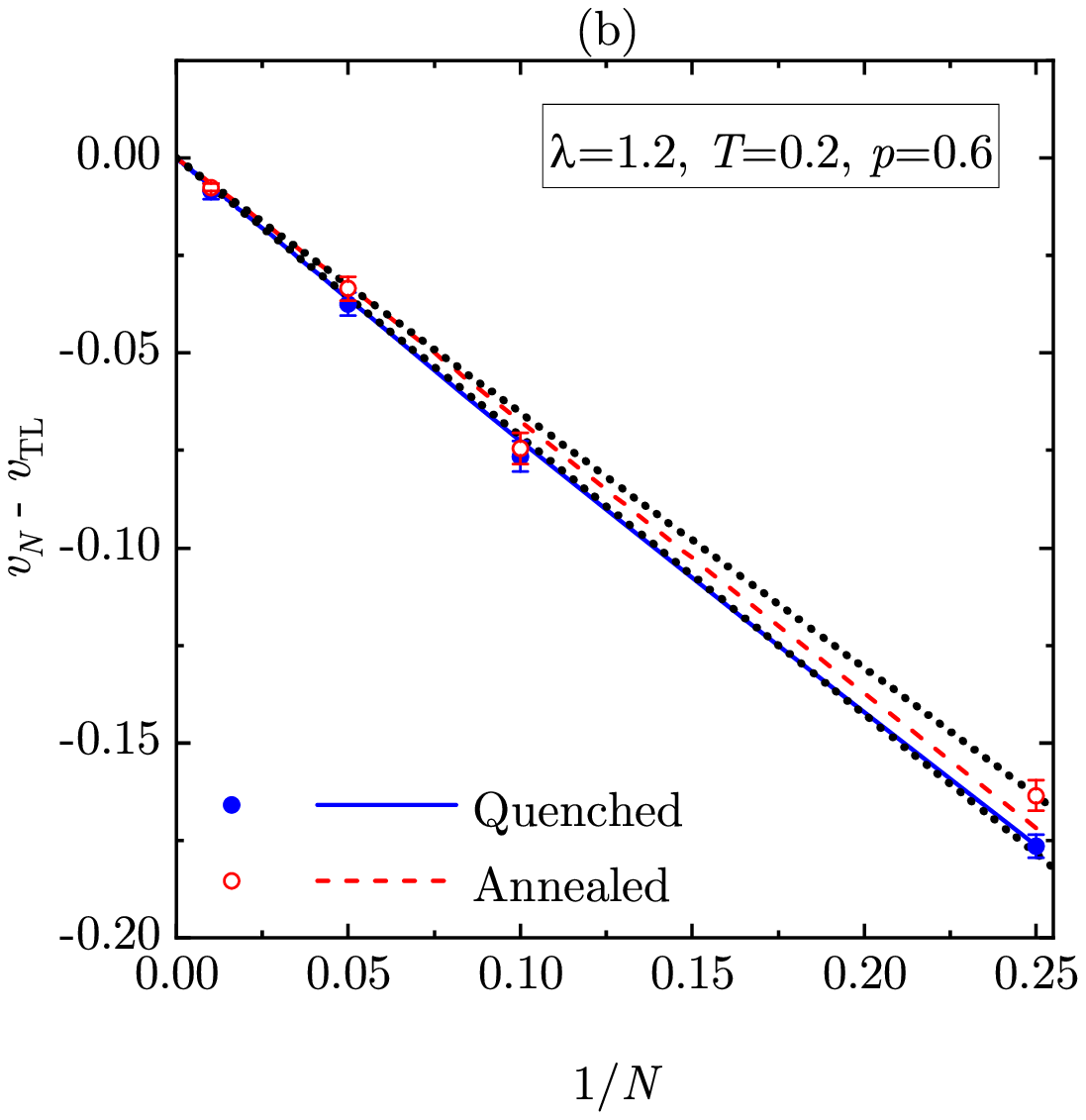}
\caption{Plot of the finite-$N$ correction $v_N-v_{\text{TL}}$ vs $1/N$ for $\lambda=1.2$, $p=0.6$, and (a) $T=1$ and (b) $T=0.2$. The filled circles and solid lines correspond to  MC simulations and exact theoretical results, respectively, for an equimolar ($x_1=x_2=\frac{1}{2}$) quenched mixture, while the open circles and dashed lines correspond to  MC simulations and exact theoretical results, respectively, for an annealed system. The dotted lines represent the exact asymptotic behaviors. Note that the asymptotic and full lines for the quenched and annealed systems are practically indistinguishable  in panel (a).}
\label{fig3}
\end{center}
\end{figure}

\begin{figure}
\begin{center}
\includegraphics[width=.5\columnwidth]{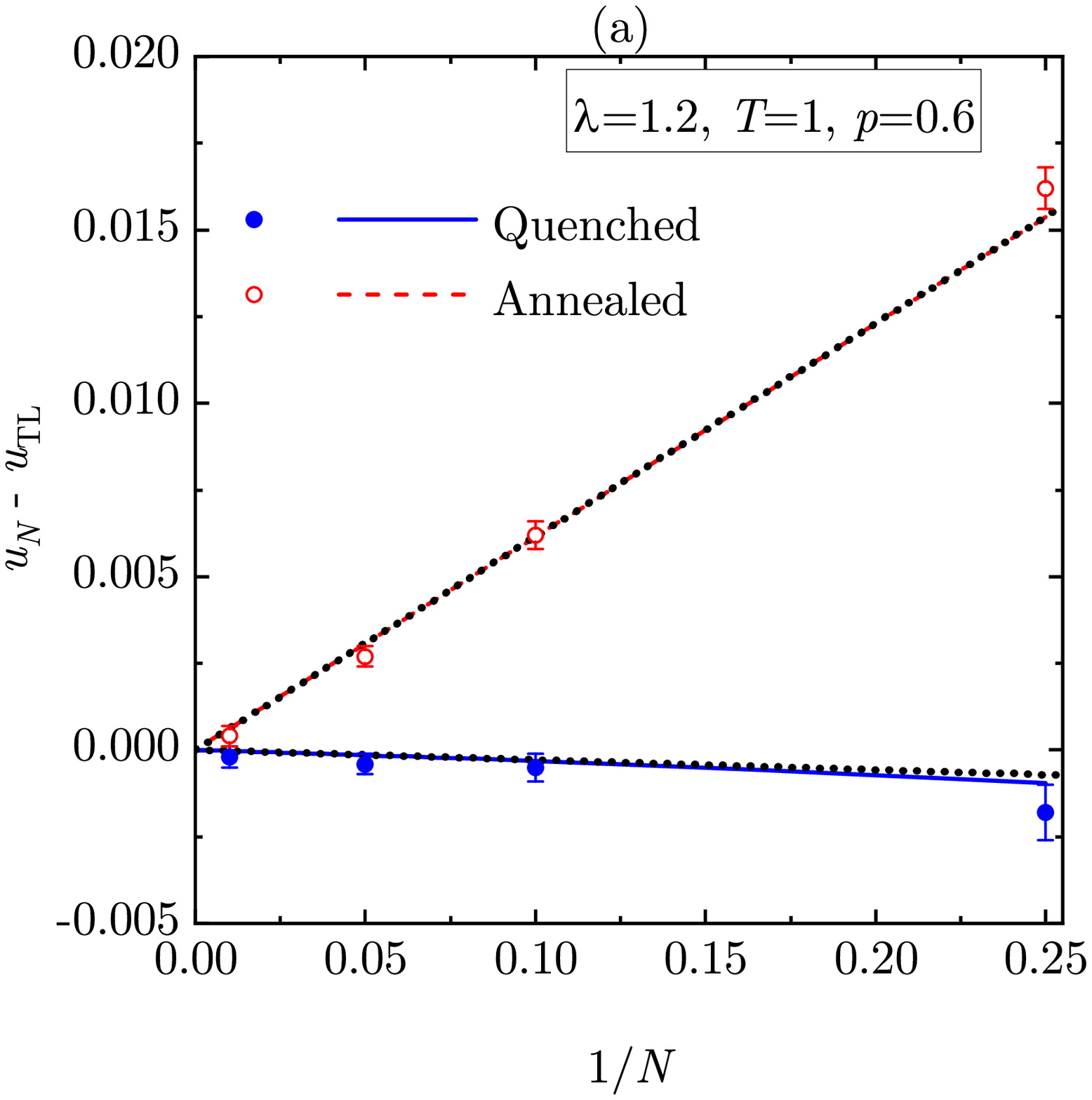}\includegraphics[width=.5\columnwidth]{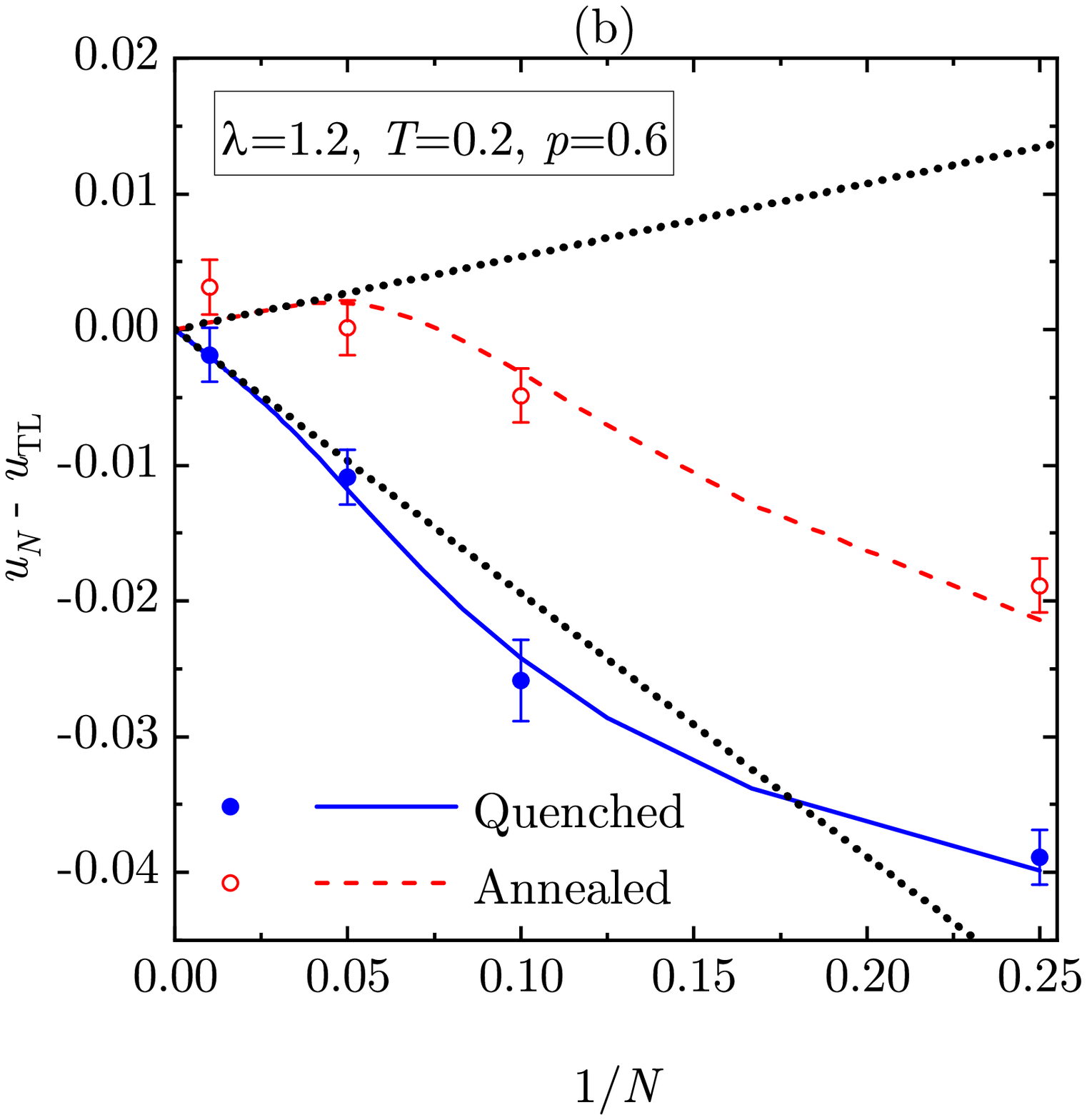}
\caption{Plot of the finite-$N$ correction $u_N-u_{\text{TL}}$ vs $1/N$  for $\lambda=1.2$, $p=0.6$, and (a) $T=1$ and (b) $T=0.2$. The filled circles and solid lines correspond to  MC simulations and exact theoretical results, respectively, for an equimolar ($x_1=x_2=\frac{1}{2}$) quenched mixture, while the open circles and dashed lines correspond to  MC simulations and exact theoretical results, respectively, for an annealed system. The dotted lines represent the exact asymptotic behaviors. Note that the asymptotic and full lines for the annealed system are practically indistinguishable  in panel (a). }
\label{fig4}
\end{center}
\end{figure}

Tables \ref{table2} and \ref{table3} give the MC results of $v_N$ and $-u^\ex_N$, respectively, for $p=0.6$, $T=1$ and $0.2$, and $N=4$, $10$, $20$, and $100$. Tables \ref{table2} and \ref{table3} also include the exact theoretical values given by equations \eref{vexN} and \eref{vexann} for $v_N$ and by equations \eref{uexN}  and  \eref{uexann} for $-u_N^\ex$.
The deviations from the TL values are displayed in figures \ref{fig3} and \ref{fig4}, which also include the asymptotic behaviors obtained from equations \eref{4b} and \eref{5b} for (equimolar) quenched systems and from equations \eref{4c} and \eref{5c} for (unbiased) annealed systems.

We can observe from tables \ref{table2} and \ref{table3} and figures \ref{fig3} and \ref{fig4} that the simulations nicely confirm our theoretical results. The differences between quenched and annealed finite-size corrections are much more important for the energy  than for the volume. In the latter case, there is a change of the sign of $v_N-v_{\text{TL}}$ when decreasing temperature from $T=1$ to $T=0.2$. Interestingly, except for the energy at low temperature ($T=0.2$), the asymptotic behaviors given by equations \eref{4b}, \eref{5b}, \eref{4c}, and \eref{5c} apply very well for any $N$, including $N=4$.

\begin{table}
\caption{\label{table2}
Values of the average volume (length) per particle, $v_N$,  in equimolar quenched mixtures and in annealed systems for $N=4$, $10$, $20$, and $100$. In all the cases, $\lambda=1.2$ and $p=0.6$. The TL values are $v_{\text{TL}}=2.6000$ and $1.2265$ at $T=1$ and $0.2$, respectively.}
\begin{indented}
\item[]
\hspace{-1.5cm}
\begin{tabular}{@{}cccccccccccc}
\br
&\multicolumn{5}{c}{$T=1$}&&\multicolumn{5}{c}{$T=0.2$}\\
\cline{2-6} \cline{8-12}
&\multicolumn{2}{c}{Quenched}&&\multicolumn{2}{c}{Annealed}&&\multicolumn{2}{c}{Quenched}&&\multicolumn{2}{c}{Annealed}\\
\cline{2-3}\cline{5-6}\cline{8-9}\cline{11-12}
$N$&Exact&MC&&Exact&MC&&Exact&MC&&Exact&MC\\
\mr
$4$&$2.7658$&$2.77(2) $&&$2.7819$&$2.80(2)$&&$1.0502$&$1.050(3)$&&$1.0547$&$1.063(4)$ \\
$10$&$2.6664$&$2.68(1)$&&$2.6728$&$2.69(1)$&&$1.1540$&$1.150(4)$&&$1.1591$&$1.152(4)$ \\
$20$&$2.6332$&$2.646(5)$&&$2.6364$&$2.647(5)$&&$1.1903$&$1.189(3)$&&$1.1936$&$1.193(3)$ \\
$100$&$2.6067$&$2.612(8)$&&$2.6073$&$2.623(8)$&&$1.2194$&$1.218(2)$&&$1.2200$&$1.219(1)$ \\
\br
\end{tabular}
\end{indented}
\end{table}

\begin{table}
\caption{\label{table3}
Absolute values of the excess energy per particle, $-u^\ex_N$,  in equimolar quenched mixtures and in annealed systems for $N=4$, $10$, $20$, and $100$. In all the cases, $\lambda=1.2$ and $p=0.6$. The TL values are $-u^\ex_{\text{TL}}=0.06720$ and $0.4421$ at $T=1$ and $0.2$, respectively.}
\begin{indented}
\item[]
\hspace{-2cm}
\begin{tabular}{@{}cccccccccccc}
\br
&\multicolumn{5}{c}{$T=1$}&&\multicolumn{5}{c}{$T=0.2$}\\
\cline{2-6} \cline{8-12}
&\multicolumn{2}{c}{Quenched}&&\multicolumn{2}{c}{Annealed}&&\multicolumn{2}{c}{Quenched}&&\multicolumn{2}{c}{Annealed}\\
\cline{2-3}\cline{5-6}\cline{8-9}\cline{11-12}
$N$&Exact&MC&&Exact&MC&&Exact&MC&&Exact&MC\\
\mr
$4$&$0.06815$&$0.0690(8)$&&$0.05183$&$0.0510(6)$&&$0.4820$&$0.481(2)$&&$0.4635$&$0.461(2)$ \\
$10$&$0.06752$&$0.0677(4)$&&$0.06105$&$0.0610(4)$&&$0.4664$&$0.468(3)$&&$0.4453$&$0.447(2)$ \\
$20$&$0.06735$&$0.0676(3)$&&$0.06412$&$0.0645(3)$&&$0.4539$&$0.453(2)$&&$0.4402$&$0.442(2)$ \\
$100$&$0.06723$&$0.0674(3)$&&$0.06658$&$0.0668(3)$&&$0.4441$&$0.444(2)$&&$0.4416$&$0.439(2)$ \\
\br
\end{tabular}
\end{indented}
\end{table}

\subsection{Biased annealed systems}

\begin{figure}
\begin{center}
\includegraphics[width=.5\columnwidth]{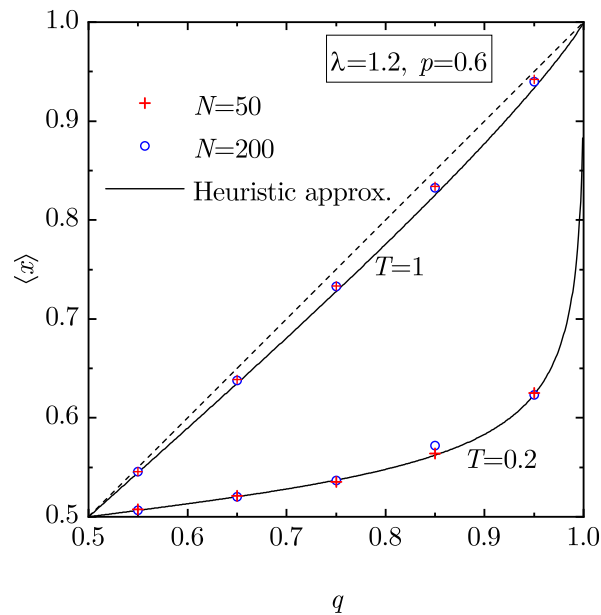}
\caption{Plot of the average mole fraction $\langle x\rangle$ vs $q$ for biased annealed systems, as obtained from MC simulations with $N=50$ and $200$  for $\lambda=1.2$, $p=0.6$, and  $T=1$ and  $0.2$. The size of the symbols is larger than the error bars. The solid lines represent the simple heuristic approximation given by the solution to equation \eref{x_q} with $a=10$, while  the straight dashed line is the reference $\langle x\rangle =q$. }
\label{fig5}
\end{center}
\end{figure}

\begin{figure}
\begin{center}
\includegraphics[width=.5\columnwidth]{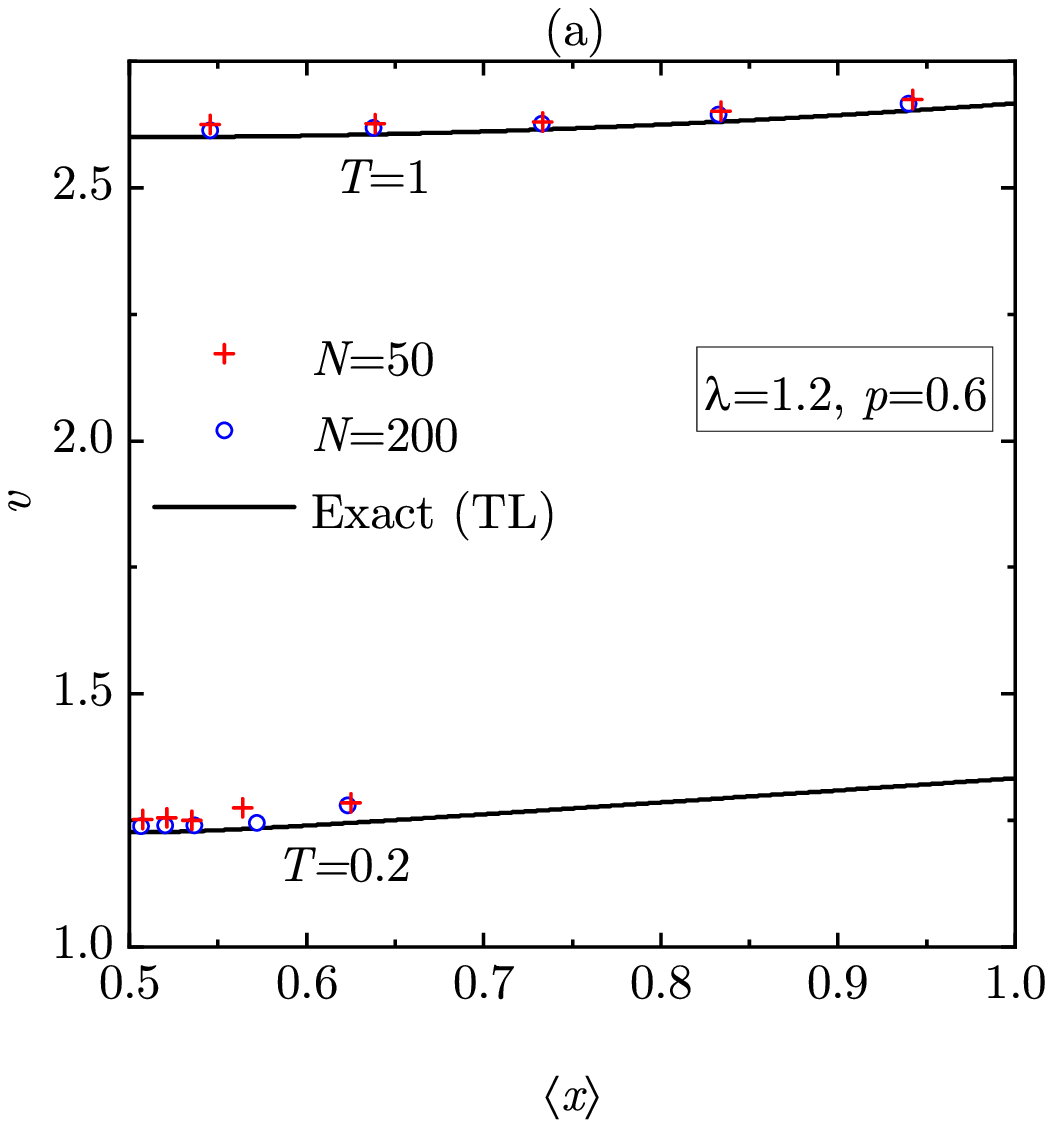}\includegraphics[width=.5\columnwidth]{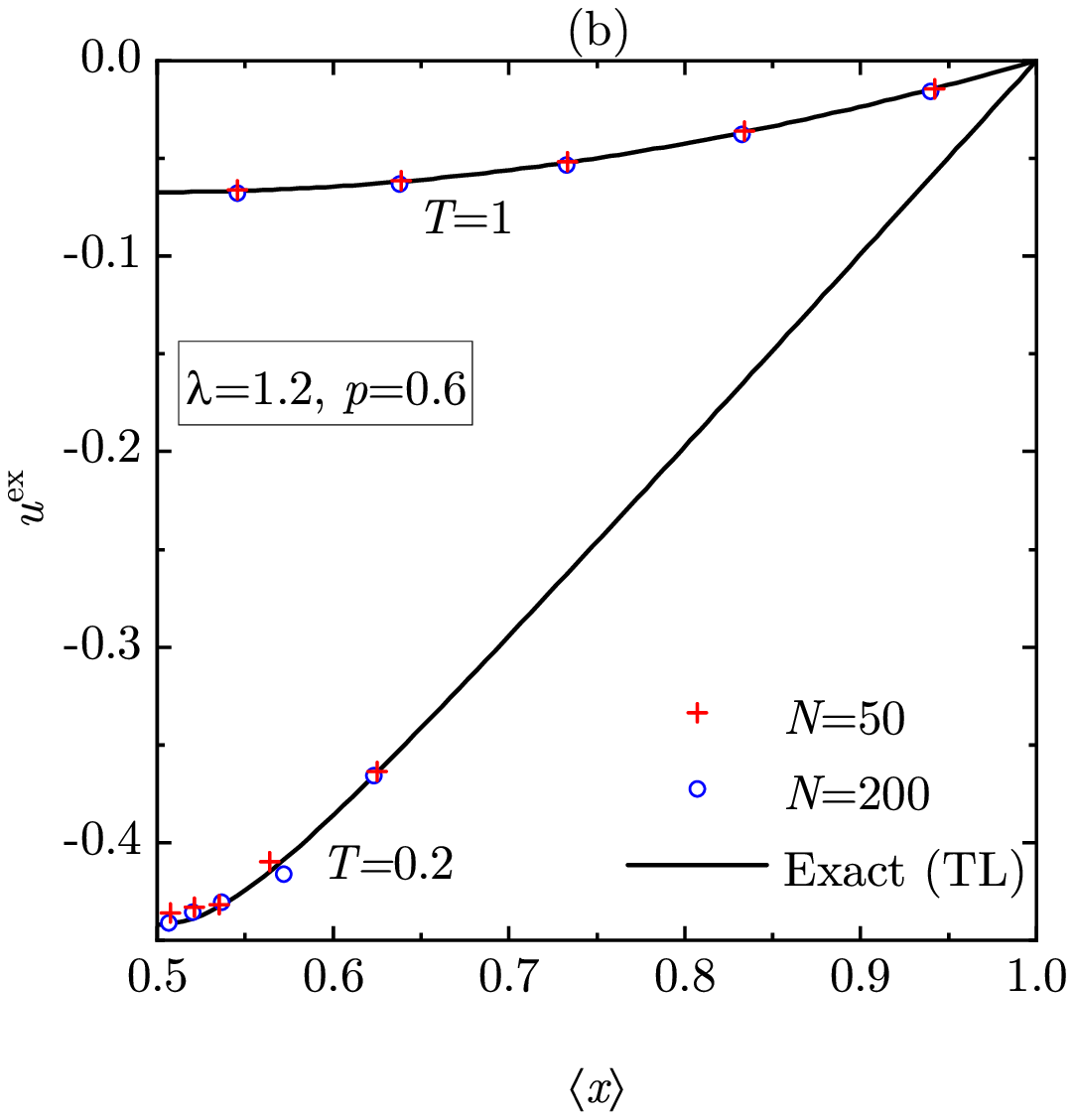}
\caption{Plot of (a) the volume $v$ and (b) the excess internal energy $u^\ex$ vs the average mole fraction $\langle x\rangle$  for biased annealed systems, as obtained from MC simulations with $N=50$ and $200$  for $\lambda=1.2$, $p=0.6$, and  $T=1$ and  $0.2$. The size of the symbols is larger than the error bars. The lines represent the exact theoretical results in the TL.}
\label{fig6}
\end{center}
\end{figure}

The MC simulations for annealed systems presented above are \emph{unbiased} in the sense that, even though the identities of the particles are not fixed and thus the mole fraction $x_1$ is a fluctuating quantity, no preference to either spin orientation is imposed, so that $\langle x_1\rangle=\frac{1}{2}$. As a consequence,  the unbiased annealed results become equivalent to the equimolar quenched ones in the TL.

On the other hand, it is possible to carry out \emph{biased} annealed simulations by introducing a parameter $q\neq \frac{1}{2}$ which favors one of the two possible spin orientations (see \ref{appMC}). As observed in reference \cite{MS20}, the average value $\langle x_1\rangle\equiv \langle x\rangle$ does not coincide with $q$, but a natural question arises as to whether or not the inequality $\langle x\rangle\neq q$ is a finite-size artifact.

To address that question, we have performed MC simulations for biased annealed systems with $q=0.55$, $0.65$, $0.75$, $0.85$, and $0.95$. As before, we have fixed $\lambda=1.2$, $p=0.6$, and temperatures $T=1$ and $0.2$. As for the number of particles, the values $N=50$ and $200$ have been chosen. The results are displayed in figure \ref{fig5}, which shows that the data with $N=50$ and $200$ practically coincide. Therefore, the property  $\langle x\rangle\neq q$ (actually, $\frac{1}{2}\leq\langle x\rangle\leq q$ or  $q\leq \langle x\rangle \leq\frac{1}{2}$) and the dependence  $\langle x\rangle(q)$ are robust with respect to $N$ and must hold in the TL. While the derivation of the exact function $\langle x\rangle(q)$ seems to be rather involved and lies outside of the scope of this work, we have constructed a simple heuristic approximation in \ref{appD}. Figure \ref{fig5} shows that  equation \eref{x_q} with $a=10$ displays an excellent agreement with the MC data.

In the MC simulations for biased annealed systems we have also evaluated the specific volume ($v$) and the excess internal energy per particle ($u^\ex$). Once the robustness of the relationship $\langle x\rangle(q)$ has been checked, one can take $q$ as a parameter and plot $v$ and $u^\ex$ as functions of the mole fraction $\langle x \rangle$. This is done in figure \ref{fig6}. While in the case $T=1$ the mapped range is
$0.55\lesssim \langle x\rangle \lesssim 0.94$, the range shrinks to $0.51\lesssim \langle x\rangle \lesssim 0.63$ if $T=0.2$.
Again, a very weak influence of $N$ is observed. As a matter of fact, comparison with the exact theoretical results for non-equimolar mixtures in the TL [see equations \eref{vTL} and \eref{uTL}] presents a very good agreement. It is worth mentioning that $v$ exhibits a rather weak dependence on the mole fraction, with a local minimum at $\langle x\rangle=\frac{1}{2}$. On the other hand, the excess energy $u^\ex$ is much more sensitive to $\langle x\rangle$, vanishing at $\langle x\rangle=0$ and $\langle x\rangle=1$, as expected.

\section{Conclusions}
\label{sec6}

This paper has focused on the study of finite-size effects on the thermodynamic quantities of Janus fluids confined to one-dimensional configurations. Two classes of systems (quenched and annealed) have been considered. In the quenched case, the fraction $x_i$ of particles with a particular face (or spin)  orientation is kept fixed. On the other hand, particles can flip their orientations in annealed systems, so that the mole fraction $x_i$ fluctuates around a value $\langle x_i\rangle =\frac{1}{2}$ (unbiased case, $q_i=\frac{1}{2}$) or $\langle x_i\rangle \neq \frac{1}{2}$ (biased case, $q_i\neq\frac{1}{2}$).

Our study allows us to answer affirmatively the four questions initially posed in section \ref{sec1}:
\begin{enumerate}[i]
\item
Can the exact derivation of the Gibbs free energy in the TL ($N\to\infty$) be extended to quenched and/or annealed finite-$N$ systems?

By working on the isothermal-isobaric ensemble with open boundary conditions, we have been able to derive exactly the configuration integral (and hence the Gibbs free energy, the specific volume, and the internal energy) for quenched systems with arbitrary values of number of particles $N$, mole fraction $x_1$, temperature $T$, pressure $p$, and nearest-neighbor interactions $\phi_{11}$ and $\phi_{12}$. The results are summarized by equations \eref{QN_f}--\eref{uexN}.

The exact results  for quenched systems are next exploited to get the finite-size quantities for unbiased annealed systems, as given by equations \eref{QN_ann}--\eref{uexann}.

\item
Does the quenched$\leftrightarrow$annealed equivalence break down at finite $N$?

The exact results referred to in the previous point apply to any finite $N$. An interesting problem consists in taking the limit $N\to\infty$ in order to obtain well-defined expressions for the thermodynamic quantities in the TL, as well as the first $N^{-1}$-correction. This is done in \ref{appA} and \ref{appC}, the correction results being given by equations \eref{gNapprox_a}, \eref{4}, and \eref{5} for the quenched case and by equations \eref{gNequi_b}--\eref{5c} for the unbiased annealed case.

The quenched quantities in the TL are provided by equations \eref{gTLa}, \eref{vTL}, and \eref{uTL}. As proved in \ref{appB}, equation \eref{gTLa} is equivalent to (but more compact than) the Gibbs free energy derived in reference \cite{MS20} from a completely different method. While in reference \cite{MS20} the thermodynamic results were derived firectly in the TL from the structural correlation functions, here they have been derived by carefully taking the limit $N\to\infty$ from the configuration integral. The equivalence between both routes reinforces the exact character of the results.

The results for equimolar quenched systems  and those for unbiased annealed systems agree in the TL [equations \eref{gTLequi}--\eref{uTLequi}], but they differ in the first $N^{-1}$-correction [compare equations \eref{gNequi}--\eref{5b} with equations \eref{gNequi_b}--\eref{5c}]. Therefore, the quenched$\leftrightarrow$annealed equivalence does break down at finite $N$.

\item
Can those theoretical predictions be validated by MC simulations?

The conclusions summarized by the two preceding points apply to any choice of the interaction potentials $\phi_{11}$ and $\phi_{12}$. In order to validate them by simulations, we have specialized to the Kern--Frenkel model \cite{KF03}, as defined by equation \eref{KF}. MC results have been measured for a well range $\lambda=1.2$, a common pressure $p=0.6$, two temperatures ($T=1$ and $0.2$), and four values of the number of particles ($N=4$, $10$, $20$, and $100$). As shown by figures \ref{fig3} and \ref{fig4}, the agreement is very good. Interestingly, except for the case of the internal energy at $T=0.2$, the deviations from the TL values closely follow the $N^{-1}$ rule even for system sizes  as small as $N=4$.

\item
Is the dependence of the average mole fraction $\langle x\rangle$ on the probability $q$  robust with respect to $N$ in annealed MC simulations for biased situations ($q\neq \frac{1}{2}$)?

The finite-size corrections mentioned above for annealed systems apply to unbiased situations. In particular, in each MC step an attempt to assign the orientation identity $i=1$  to a given particle is carried out with a probability $q=\frac{1}{2}$, what results in an average mole fraction $\langle x\rangle=\frac{1}{2}$. The procedure can be extended in a straightforward way to a biased choice $q\neq \frac{1}{2}$, which gives rise to $\langle x\rangle \neq\frac{1}{2}$. The naive expectation would be $\langle x\rangle=q$, but preliminary results in reference \cite{MS20} showed that either $\frac{1}{2}<\langle x\rangle<q$ or $\frac{1}{2}>\langle x\rangle>q$, depending on whether $q>\frac{1}{2}$ or $q<\frac{1}{2}$, respectively. One might reasonable wonder whether the property $\langle x\rangle\neq q$ is a finite-size effect that would disappear in the TL.

However, our MC results provide strong evidence about the robustness of the inequality $\langle x\rangle\neq q$ and the dependence of $\langle x\rangle$  on $q$ (see figure \ref{fig5}). This can be qualitatively explained as follows. In the quenched case, the configuration integral presents a peaked local maximum at $N_1=N/2$, i.e., $x=\frac{1}{2}$, as can be seen from equations \eref{QN_f}, \eref{Xi_N1N2_c}, and \eref{E1}. For annealed systems, this competes against a weight function $w_N(x)$ exhibiting a peaked local maximum at $x=q$. The annealed  probability density $P_N(x)$ is proportional to the product of both functions and then it has a peaked maximum at an intermediate value $x=\langle x \rangle$. Based on these arguments, a heuristic approach has been put forward in \ref{appD}. Its theoretical predictions (with a single fitting parameter $a=10$ independent of $T$ and $q$) agree excellently well with MC simulations, as figure \ref{fig5} shows.

As a bonus of the biased annealed simulations, and given the weak influence of $N$ observed in figure \ref{fig5}, we have compared the measured MC values of volume and energy  with the theoretical exact results in the TL as functions of the mole fraction. The results displayed by figure \ref{fig6} show again an excellent agreement.

\end{enumerate}

To put our findings in a proper context, some of their limitations should be remarked. First, the theoretical results have been obtained for open boundary conditions [$\omega=0$ in equation \eref{N1}]. As shown by equation \eref{QN_b}, application of periodic boundary conditions ($\omega=1$) significantly hampers the quest for an exact treatment at finite $N$. While the choice of the boundary conditions (open or periodic) becomes irrelevant in the TL, finite-size effects are affected by such a choice.

A second limitation arises from the use of the isothermal-isobaric ensemble rather than the standard canonical ensemble. Of course, the partition function and its associated configuration integral can be formally written in the canonical ensemble [consider equation \eref{QN} with the integration over $L$ removed], but then it is much more difficult to reduce the problem to a purely combinatorial one at finite $N$, as happens, however,  with equations \eref{QN_c}--\eref{QN_e}. One might believe that it would be possible to get the finite-size Helmholtz free energy from the finite-size Gibbs free energy derived here by means of the conventional Legendre transformation. However, this transformation is justified in the TL only and washes out finite-size effects, as we have checked by comparison with canonical MC simulations (not shown).

Third, we have not addressed in the present paper the problem of deriving the exact relationship between $\langle x\rangle$ and $q$ in biased annealed systems, even in the TL. The theoretical approach in \ref{appD} is heuristic and depends upon a parameter $a$ whose value must be obtained by a fitting procedure. It would be very interesting to analyze in detail the random walk represented by the annealed MC simulations and derive the dependence $\langle x\rangle (q)$, at least in the TL. However, this goal is outside of the scope of the present work.

The last limitation refers to the choice of the one-dimensional geometry itself. Of course, two- and three-dimensional systems are much more realistic, but the one-dimensional setting, apart from being applicable to single-file confinement situations, has the enormous advantage of allowing for the derivation of nontrivial exact results. For instance, we have explicitly shown in a clean way that the first corrections to the TL values are of order $N^{-1}$, as usually assumed in the literature to get rid of finite-size effects and extrapolate the simulation data to the TL. Moreover, exact results are utterly important to test simulation methods and/or theoretical approaches that can then be extended to scenarios where exact solutions are absent.

\ack

A S acknowledges financial support from Grant PID2020-112936GB-I00 funded by MCIN/AEI/10.13039/501100011033, and from Grants IB20079 and GR18079 funded by Junta de Extremadura (Spain) and by ERDF: A way of making Europe.

\appendix
\section{Function $\Xi_{N_1,N_2}$ for large $N$}
\label{appA}
In this appendix, we prove that the function $\Xi_{N_1,N_2}$ defined in equations \eref{QN_f} and \eref{xi} reduces to equation \eref{Xi_N1N2_c} in the limit $N\to \infty$.

First, application of the Stirling approximation $x!\approx \sqrt{2\pi x}(x/\rme)^x$ yields
\begin{equation}
\label{1}
\xi_{N_1,N_2}(n=Ny)\approx \exp\left[N\psi(y)\right],\quad \psi(y)=\psi_0(y)+N^{-1}\psi_{1}(y),
\end{equation}
where
\begin{equation}
\label{psi0}
\psi_0(y)=-x_1\ln\left(1-\frac{y}{x_1}\right)-x_2\ln\left(1-\frac{y}{x_2}\right)+y\ln\frac{(x_1-y)(x_2-y)}{y^2(1-R)},
\end{equation}
\begin{equation}
\label{psi1}
\psi_{1}(y)=-\ln\left[2\pi Ny\sqrt{\left(1-\frac{y}{x_1}\right)\left(1-\frac{y}{x_1}\right)}\right].
\end{equation}
Equating to zero the first derivative of $\psi(y)$ with respect to $y$, one can find that the maximum value of $\psi(y)$ corresponds to
\begin{equation}
y_{\max}\approx y_0+N^{-1} y_1,
\end{equation}
where
\begin{equation}
\label{y0}
y_0= \frac{1-\sqrt{1-4x_1 x_2R}}{2R},\quad
y_1=-\frac{1+(4y_0-3)y_0/2x_1x_2}{2-y_0/x_1x_2}.
\end{equation}
Note that $\psi_0'(y_0)=0$ and $y_{1}=-\psi_{1}'(y_0)/\psi_0''(y_0)$, where the second derivative of the $\psi_0(y)$ is
\begin{equation}
\label{psi0''}
\psi_0''(y)=-\frac{2-y/x_1x_2}{y(1-y/x_1)(1-y/x_2)}.
\end{equation}
Note also that the last term on the right-hand side of equation \eref{psi0} vanishes at $y=y_0$, so that $\bar{\psi}_0\equiv \psi_0(y_0)$ is given by equation \eref{psi00}

As a second step, let us expand $\psi(y)$ around $y=y_{\max}$ to get
\begin{equation}
\label{2}
\psi(y)\approx \psi(y_{\max})+\frac{\psi''(y_{\max})}{2}\left(y-y_{\max}\right)^2.
\end{equation}
Next, we replace the sum in $\Xi_{N_1,N_2}$ by an integral:
\begin{eqnarray}
\label{Xi_N1N2}
\Xi_{N_1,N_2}&\approx& N\int_{-\infty}^\infty \rmd y\, \xi_{N_1,N_2}(Ny)\nonumber\\
&\approx& N \rme^{N\psi(y_{\max})}\int_{-\infty}^\infty \rmd y\, e^{\frac{N\psi''(y_{\max})}{2}\left(y-y_{\max}\right)^2}\nonumber\\
&=&N \rme^{N\psi(y_{\max})}\sqrt{\frac{2\pi}{-N\psi''(y_{\max})}},
\end{eqnarray}
where in the second step use has been made of equation \eref{2}.
Finally, taking into account that $\psi(y_{\max})\approx \psi_0(y_0)+N^{-1}\psi_1(y_0)$ and $\psi''(y_{\max})\approx \psi_0''(y_0)$, equation \eref{Xi_N1N2} becomes
\begin{equation}
\label{Xi_N1N2_b}
\Xi_{N_1,N_2}\approx N \rme^{N\psi_0(y_{0})+\psi_1(y_0)}\sqrt{\frac{2\pi}{-N\psi_0''(y_{0})}}.
\end{equation}
Insertion of equations \eref{psi1} and \eref{psi0''} into equation \eref{Xi_N1N2_b} yields equation \eref{Xi_N1N2_c}.

\section{Proof of equation \eref{proof}}
\label{appB}
While $\bar{\psi}_0$ is expressed in terms of $y_0$ [see equation \eref{psi00}], the right-hand side of equation \eref{proof} is expressed in terms of $R$. The latter quantity is related to $y_0$ by the identities
\begin{equation}
\fl
R=\frac{y_0-x_1 x_2}{y_0^2},\quad \sqrt{1-4x_1 x_2 R}=\frac{2x_1 x_2}{y_0}-1,\quad \sqrt{1-R}=\frac{\sqrt{(x_1-y_0)(x_2-y_0)}}{y_0},
\end{equation}
\begin{equation}
\label{4x1x2}
\frac{1+\sqrt{1-4x_1x_2R}}{2\sqrt{1-R}}=\frac{x_1 x_2}{\sqrt{(x_1-y_0)(x_2-y_0)}},
\end{equation}
\begin{equation}
\label{x1-x2}
\frac{|x_1-x_2|+\sqrt{1-4x_1x_2R}}{(|x_1-x_2|+1)\sqrt{1-R}}=\frac{x_2}{x_1}\sqrt{\frac{x_1-y_0}{x_2-y_0}},
\end{equation}
where, without loss of generality, we have assumed $x_1\geq x_2$ in equation \eref{x1-x2}.

The right-hand side of equation \eref{proof} can be rewritten as
\begin{eqnarray}
\text{r.h.s.}&=&-x_1 \ln \left[x_1\frac{2\sqrt{1-R}}{1+\sqrt{1-4x_1x_2R}}\frac{x_1-x_2+\sqrt{1-4x_1x_2R}}{(x_1-x_2+1)\sqrt{1-R}}\right]\nonumber\\
&&-x_2\ln \left[x_2\frac{2\sqrt{1-R}}{1+\sqrt{1-4x_1x_2R}}\frac{(x_1-x_2+1)\sqrt{1-R}}{x_1-x_2+\sqrt{1-4x_1x_2R}}\right]\nonumber\\
&=&-x_1\ln\left(1-\frac{y_0}{x_1}\right)-x_2\ln\left(1-\frac{y_0}{x_2}\right),
\end{eqnarray}
where we have made use of equations \eref{4x1x2} and \eref{x1-x2}. Comparison with equation \eref{psi00} closes the proof of equation \eref{proof}.

\section{Function $\Xi_{N}$ for large $N$}
\label{appC}
The method is analogous to the one followed in appendix \ref{appA}.
The quantities $\bar{\psi}_0$ and $y_0$ defined in equation \eref{psi00} are functions of the mole fraction $x_1$. It can be checked that $\bar{\psi}_0$ presents a maximum at $x_1=\frac{1}{2}$. Expanding in powers of $x_1-\frac{1}{2}$,
\begin{equation}
\label{psi0x1}
\bar{\psi}_0\approx \ln\left(1+\frac{1}{\sqrt{1-R}}\right)-\frac{2}{\sqrt{1-R}}\left(x_1-\frac{1}{2}\right)^2.
\end{equation}
Combination of equations \eref{Xi_N1N2_c} and \eref{psi0x1} yields
\begin{equation}
\label{C.2}
\Xi_{N_1,N_2}\approx \left(1+\frac{1}{\sqrt{1-R}}\right)^N\frac{1+\sqrt{1-R}}{\sqrt{2\pi N\sqrt{1-R}}}\rme^{-2N\left(x_1-\frac{1}{2}\right)^2/\sqrt{1-R}}.
\end{equation}

As a second step, for large $N$ the summation of $\Xi_{N_1,N_2}$ over $N_1$ can be approximated by an integral over $x_1$:
\begin{equation}
\fl
\sum_{N_1=0}^{N}\Xi_{N_1,N_2}\approx \left(1+\frac{1}{\sqrt{1-R}}\right)^N\frac{1+\sqrt{1-R}}{\sqrt{2\pi N\sqrt{1-R}}}N\int_{-\infty}^\infty \rmd x_1\,\rme^{-2N\left(x_1-\frac{1}{2}\right)^2/\sqrt{1-R}}.
\end{equation}
This finally gives equation \eref{sumXi}.

\section{Technical details of the MC simulations}
\label{appMC}

Since our exact finite-size results are found in the isothermal-isobaric ensemble and
the Legendre transform `washes out' the finite-size effects, we found it necessary to
perform our numerical experiments also in the isothermal-isobaric ensemble \cite{FS02}.
Moreover, in order to find agreement with our theoretical exact results,  open boundary conditions were used. Of course, only in the TL open and periodic
boundary conditions become equivalent.

We performed two kinds of MC experiments, which we label as MCa and MCq for annealed and quenched systems, respectively.

The MCa transition rule consists of single particle MC moves (one MC step), which are
the combination of a particle position displacement
$x_\alpha\to x_\alpha+(2\eta-1)\delta$, where $\eta$ is a pseudo-random number in
$[0,1]$ and $\delta<\sigma$ is the maximum displacement (to be kept fixed during
the whole simulation  to preserve  detailed balance) and a particle assignment to
species $i=1,2$ with probability $q_i$ (where $q_1=q$ and $q_2=1-q$). Open boundary conditions were enforced by generating a new
position until it falls inside the segment $x_\alpha\in[-L/2,L/2]$. According to the
Metropolis algorithm \cite{MRRTT53,KW08} the move is accepted with
probability $\rme^{-\beta\Delta\Phi_N}$, $\Delta\Phi_N$ being the change in potential energy
due to the combined move. This would be enough in the canonical ensemble, while in the
isothermal-isobaric ensemble we also need to perform a volume move. The latter is computationally
the most expensive one, since it requires a full energy calculation at each attempt and
therefore should be used with a low frequency during the run. We chose 30\% for the
frequency of the volume move in all our simulations. For the transition and
acceptance probability for this volume move, see for example reference \cite{FS02}.

In contrast to the MCa case, in the MCq simulations the particles are assigned an identity $i=1,2$ with probability $x_i=q_i$ from the start and the species assignment is never changed afterwards. The MCq transition rule consists of single particle MC moves that amount to a
particle position displacement with $\delta>\sigma$ (note that this condition may be
relieved in dimensions higher than one), which is accepted with probability
$\rme^{-\beta\Delta\Phi_N}$, $\Delta\Phi_N$ being the change in potential energy due to
the displacement.  Again, in the isothermal-isobaric ensemble we also
have the volume move \cite{FS02}.

Notice that we can obtain the same result for quenched systems by using a third simulation
strategy that we will call MCaq. The MCaq transition rule consists of single particle
MC moves that are the combination of a particle position displacement (with
$\delta>\sigma$), which is accepted with probability $\rme^{-\beta\Delta\Phi_N}$ (where
$\Delta\Phi_N$ is the change in potential energy due to the displacement only), followed by  a
particle assignment to species $i=1,2$ with probability $q_i$, which is always accepted and
therefore completely disentangled from the displacement move. As before,  we also have the volume move \cite{FS02} in the
isothermal-isobaric ensemble.

In all cases we chose $\delta$ so to have acceptance ratios as close as possible to
$\frac{1}{2}$. The equilibration time for MCa was much longer than for MCq.

Given an observable $\calo$, its statistical-mechanical average $\langle \calo\rangle$ was evaluated by averaging $\calo$ over a sufficiently large number of MC configurations after a sufficiently long equilibration time. The measured observables were the mole fraction $x=N^{-1}\sum_{\alpha=1}^N\delta_{s_\alpha,1}$, the specific volume (or reciprocal density) $v=L/N$, and the excess
internal energy per particle $u^\ex=\Phi_N/N$.

The statistical error on $\langle\calo\rangle$ is as usual given by
$\sigma_{\langle\calo\rangle}=\sqrt{\sigma_\calo^2\tau_\calo/M}$, where
$M$ is the number of MC steps, $\sigma_\calo^2$ is the intrinsic variance of $\calo$,
and $\tau_\calo$ is the correlation time for the observable $\calo$
\cite{KW08}. The latter quantity depends crucially on the transition
rule and has a minimum value equal to $1$ if one can move so far in configuration space that
successive values become uncorrelated. In general, the number of independent steps which
contribute to reducing the error bar is not $M$ but $M/\tau_\calo$. Hence, to determine the
true statistical error in the random walk, one needs to estimate the correlation time.
To do this, it is very important that the total length of the random walk be much
greater than $\tau_\calo$. Otherwise, the result and its error bar will not be reliable. In
general, there is no mathematically rigorous procedure to determine $\tau_\calo$, so that usually one
must determine it from the random walk itself. It is a good practice occasionally to carry out
very long runs to test that the results are well converged. In order to equilibrate the random walk, we generally found it necessary to
use $10^6$ MC steps at high temperature ($T=1$) and $2\times 10^7$ MC steps at low
temperature ($T=0.2$), and collect averages over $M=10^5$ MC steps.

\section{A heuristic approximation for the dependence of $\langle x\rangle$ on $q$ for biased annealed systems}
\label{appD}

From equations \eref{C.2} and \eref{sumXi}, we have that, for large $N$, the probability  that the mole fraction $x_1$ lies between $x$ and $x+\rmd x$ in the unbiased annealed system is
\begin{equation}
\label{E1}
\fl
P_N(x)\rmd x=\frac{1}{\Xi_N}\sum_{N_1=Nx}^{N(x+\rmd x)}\Xi_{N_1,N_2}
\approx\frac{N \Xi_{N x,N(1-x)}}{\Xi_N}\rmd x
\approx\frac{\rme^{-2N\left(x-\frac{1}{2}\right)^2/\sqrt{1-R}}}{\sqrt{\pi\sqrt{1-R}/2N}}\rmd x.
\end{equation}
Obviously, $\langle x\rangle=\frac{1}{2}$.

Imagine now a \emph{biased} annealed system where each value of $x=N_1/N$ is weighed with a certain function $w_N(x)$ centered around a value $x=q\neq \frac{1}{2}$.
In that case,
\begin{equation}
P_N(x)\propto w_N(x) \Xi_{N x,N(1-x)},
\end{equation}
which, for large $N$, would be extremely peaked around a value (comprised between $\frac{1}{2}$ and $q$) that coincides with the average $\langle x\rangle=\int_0^1\rmd x\,xP_N(x)$. Thus, the value $\langle x\rangle$ can be determined as the solution to the equation
\begin{eqnarray}
\label{E3}
0&=&\frac{\partial}{\partial x}\lim_{N\to\infty} N^{-1}\left[\ln w_N(x)+\ln \Xi_{N x,N(1-x)}\right]\nonumber\\
&=&\frac{\partial}{\partial x}\lim_{N\to\infty} N^{-1}\ln w_N(x)+\frac{\partial\bar{\psi}_0}{\partial x},
\end{eqnarray}
where in the second step we have made use of equation \eref{Xi_N1N2_c}. Note that here, in contrast to equation \eref{E1},  we need to take into account the full dependence of $\bar{\psi}_0$ on $x$ because the solution to equation  \eref{E3} is  not, in general, close to $\frac{1}{2}$. According to equation \eref{psi00},
\begin{equation}
\fl
\label{E4}
\frac{\partial\bar{\psi}_0}{\partial x}=-\ln\left[1-\frac{1-\sqrt{1-4x(1-x)R}}{2xR}\right]+\ln\left[1-\frac{1-\sqrt{1-4x(1-x)R}}{2(1-x)R}\right].
\end{equation}

The simplest choice for the weight function $w_N(x)$ is the binomial distribution $w_N(x)=\binom{N_{\text{eff}}}{N_{\text{eff}} x}q^{N_{\text{eff}}x}(1-q)^{N_{\text{eff}}(1-x)}$, where $N_{\text{eff}}\equiv N b$, $b$ being an \emph{effective} factor accounting for the expected dependence of $w_N(x)$ on the thermodynamic state ($T$ and $p$). In that case,
\begin{eqnarray}
\lim_{N\to\infty} N^{-1}\ln w_N(x)&=&b\left[x\ln\frac{q}{x}+(1-x)\ln\frac{1-q}{1-x}\right],\\
\frac{\partial}{\partial x}\lim_{N\to\infty} N^{-1}\ln w_N(x)&=&b\ln \frac{q(1-x)}{x(1-q)}.
\end{eqnarray}
Therefore, equation \eref{E3} becomes
\begin{eqnarray}
\label{x_q}
\fl
0&=&-\ln\left[1-\frac{1-\sqrt{1-4x(1-x)R}}{2xR}\right]+\ln\left[1-\frac{1-\sqrt{1-4x(1-x)R}}{2(1-x)R}\right]
\nonumber\\
\fl
&& +a\sqrt{1-R}\ln \frac{q(1-x)}{x(1-q)},
\end{eqnarray}
where we have taken $b=a\sqrt{1-R}$, $a$ being a constant to be empirically determined. A simple and yet optimal value is $a=10$.

\section*{References}
\bibliographystyle{iopart-num-long}


\begin{thebibliography}{10}
\expandafter\ifx\csname url\endcsname\relax
  \def\url#1{{\tt #1}}\fi
\expandafter\ifx\csname urlprefix\endcsname\relax\def\urlprefix{URL }\fi
\providecommand{\eprint}[2][arXiv]{#1:\linebreak[0]#2}

\bibitem{RML05}
Roh K~H, Martin D~C and Lahann J 2005 Biphasic {J}anus particles with nanoscale
  anisotropy {\em Nature Mater.\/} {\bf 4} 759--763

\bibitem{WLZL08}
Wang B, Li B, Zhao B and Li C~Y 2008 Amphiphilic {J}anus gold nanoparticles via
  combining ``solid-state grafting-to'' and ``grafting-from'' methods {\em J.
  Am. Chem. Soc.\/} {\bf 130} 11594--11595

\bibitem{WM13}
Walther A and M\"uller A~H~E 2013 Janus particles: Synthesis, self-assembly,
  physical properties, and applications {\em Chem. Rev.\/} {\bf 113} 5194--5261

\bibitem{BF01}
Binks B~P and Fletcher P~D~I 2001 Particles adsorbed at the oil-water
  interface: {A} theoretical comparison between spheres of uniform wettability
  and ``{J}anus'' particles {\em Langmuir\/} {\bf 17} 4708--4710

\bibitem{SGP09}
Sciortino F, Giacometti A and Pastore G 2009 Phase diagram of {J}anus particles
  {\em Phys. Rev. Lett.\/} {\bf 103} {237}{801}

\bibitem{YHHD10}
Yuet K~P, Hwang D~K, Haghgooie R and Doyle P~S 2010 Multifunctional
  superparamagnetic {J}anus particles {\em Langmuir\/} {\bf 26} 4281--4287

\bibitem{F13}
Fantoni R 2013 {\em The Janus Fluid: A Theoretical Perspective\/} vol 923 (New
  York: Springer)

\bibitem{OTSM15}
Onishi S, Tokuda M, Suzuki T and Minami H 2015 Preparation of {J}anus particles
  with different stabilizers and formation of one-dimensional particle arrays
  {\em Langmuir\/} {\bf 31} 674--678

\bibitem{HGM34}
Herzfeld K~F and Goeppert-Mayer M 1934 On the states of aggregation {\em J.
  Chem. Phys.\/} {\bf 2} 38--44

\bibitem{T36}
Tonks L 1936 The complete equation of state of one, two and three-dimensional
  gases of hard elastic spheres {\em Phys. Rev.\/} {\bf 50} 955--963

\bibitem{N40a}
Nagamiya T 1940 Statistical mechanics of one-dimensional substances {I} {\em
  Proc. Phys.-Math. Soc. Jpn.\/} {\bf 22} 705--720

\bibitem{N40b}
Nagamiya T 1940 Statistical mechanics of one-dimensional substances {II} {\em
  Proc. Phys.-Math. Soc. Jpn.\/} {\bf 22} 1034--1047

\bibitem{T42}
Takahasi H 1942 Eine einfache methode zur behandlung der statistischen mechanik
  eindimensionaler substanzen {\em Proc. Phys.-Math. Soc. Jpn.\/} {\bf 24}
  60--62

\bibitem{vH50}
{van Hove} L 1950 Sur l'int\'egrale de configuration pour les syst\`emes de
  particules \`a une dimension {\em Physica\/} {\bf 16} 137--143

\bibitem{SZK53}
Salsburg Z~W, Zwanzig R~W and Kirkwood J~G 1953 Molecular distribution
  functions in a one-dimensional fluid {\em J. Chem. Phys.\/} {\bf 21}
  1098--1107

\bibitem{K55b}
Kikuchi R 1955 Theory of one-dimensional fluid binary mixtures {\em J. Chem.
  Phys.\/} {\bf 23} 2327--2332

\bibitem{LPZ62}
Lebowitz J~L, Percus J~K and Zucker I~J 1962 Radial distribution functions in
  crystals and fluids {\em Bull. Am. Phys. Soc.\/} {\bf 7} 415--415

\bibitem{KT68}
Katsura S and Tago Y 1968 Radial distribution function and the direct
  correlation function for one-dimensional gas with square-well potential {\em
  J. Chem. Phys.\/} {\bf 48} 4246--4251

\bibitem{LZ71}
Lebowitz J~L and Zomick D 1971 Mixtures of hard spheres with nonadditive
  diameters: {S}ome exact results and solution of {PY} equation {\em J. Chem.
  Phys.\/} {\bf 54} 3335--3346

\bibitem{P76}
Percus J~K 1976 Equilibrium state of a classical fluid of hard rods in an
  external field {\em J. Stat. Phys.\/} {\bf 15} 505--511 ISSN 1572-9613

\bibitem{P82}
Percus J~K 1982 One-dimensional classical fluid with nearest-neighbor
  interaction in arbitrary external field {\em J. Stat. Phys.\/} {\bf 28}
  67--81 ISSN 1572-9613

\bibitem{BOP87}
Borzi C, Ord G and Percus J~K 1987 The direct correlation function of a
  one-dimensional {I}sing model {\em J. Stat. Phys.\/} {\bf 46} 51--66

\bibitem{K91}
Korteweg D~T 1891 On {V}an der {W}aals's isothermal equation {\em Nature\/}
  {\bf 45} 152--154

\bibitem{R91}
{Lord Rayleigh} 1891 On the virial of a system of hard colliding bodies {\em
  Nature\/} {\bf 45} 80--82

\bibitem{HC04}
Heying M and Corti D~S 2004 The one-dimensional fully non-additive binary hard
  rod mixture: exact thermophysical properties {\em Fluid Phase Equil.\/} {\bf
  220} 85--103

\bibitem{S07}
Santos A 2007 Exact bulk correlation functions in one-dimensional nonadditive
  hard-core mixtures {\em Phys. Rev. E\/} {\bf 76} {062}{201}

\bibitem{SFG08}
Santos A, Fantoni R and Giacometti A 2008 Penetrable square-well fluids:
  {E}xact results in one dimension {\em Phys. Rev. E\/} {\bf 77} {051}{206}

\bibitem{BNS09}
Ben-Naim A and Santos A 2009 Local and global properties of mixtures in
  one-dimensional systems. {II}. {Exact} results for the {Kirkwood}--{Buff}
  integrals {\em J. Chem. Phys.\/} {\bf 131} {164}{512}

\bibitem{FGMS10}
Fantoni R, Giacometti A, Malijevsk\'y A and Santos A 2010 A numerical test of a
  high-penetrability approximation for the one-dimensional
  penetrable-square-well model {\em J. Chem. Phys.\/} {\bf 133} {024}{101}

\bibitem{F10b}
Fantoni R 2010 Non-existence of a phase transition for penetrable square wells
  in one dimension {\em J. Stat. Mech.\/}  P07030

\bibitem{S14}
Santos A 2014 Playing with marbles: {S}tructural and thermodynamic properties
  of hard-sphere systems {\em 5th Warsaw School of Statistical Physics\/} ed
  Cichocki B, Napi\'orkowski M and Piasecki J (Warsaw: Warsaw University Press)
  \href{arXiv:1310.5578}{http://arxiv.org/abs/1310.5578}

\bibitem{S16}
Santos A 2016 {\em {A Concise Course on the Theory of Classical Liquids. Basics
  and Selected Topics}\/} ({\em Lecture Notes in Physics\/} vol 923) (New York:
  Springer)

\bibitem{FS17}
Fantoni R and Santos A 2017 One-dimensional fluids with second nearest-neighbor
  interactions {\em J. Stat. Phys.\/} {\bf 169} 1171--1201

\bibitem{MS19}
Montero A~M and Santos A 2019 Triangle-well and ramp interactions in
  one-dimensional fluids: A fully analytic exact solution {\em J. Stat.
  Phys.\/} {\bf 175} 269--288

\bibitem{MS20}
Maestre M~A~G and Santos A 2020 One-dimensional janus fluids. exact solution
  and mapping from the quenched to the annealed system {\em J. Stat. Mech.\/}
  063217

\bibitem{KF03}
Kern N and Frenkel D 2003 Fluid-fluid coexistence in colloidal systems with
  short-ranged strongly directional attraction {\em J. Chem. Phys.\/} {\bf 118}
  9882--9889

\bibitem{HM13}
Hansen J~P and McDonald I~R 2013 {\em {Theory of Simple Liquids}\/} 4th ed
  (London: Academic Press)

\bibitem{FGSP11}
Fantoni R, Giacometti A, Sciortino F and Pastore G 2011 Cluster theory of janus
  particles {\em Soft Matter\/} {\bf 7} 2419--2427

\bibitem{F12}
Fantoni R 2012 A cluster theory for a janus fluid {\em Eur. Phys. J. B\/} {\bf
  85} 108

\bibitem{MFGS13}
Maestre M~A~G, Fantoni R, Giacometti A and Santos A 2013 Janus fluid with fixed
  patch orientations: {T}heory and simulations {\em J. Chem. Phys.\/} {\bf 138}
  {094}{904}

\bibitem{FGMS13}
Fantoni R, Giacometti A, Maestre M~A~G and Santos A 2013 Phase diagrams of
  {J}anus fluids with up-down constrained orientations {\em J. Chem. Phys.\/}
  {\bf 139} {174}{902}

\bibitem{FS02}
Frenkel D and Smit B 2002 {\em Understanding Molecular Simulation: From
  Algorithms to Applications\/} 2nd ed (San Diego: Academic Press)

\bibitem{MRRTT53}
Metropolis N, Rosenbluth A~W, Rosenbluth M~N, Teller A~H and Teller E 1953
  Equation of state calculations by fast computing machines {\em J. Chem.
  Phys.\/} {\bf 21} 1087--1092

\bibitem{KW08}
Kalos M~H and Whitlock P~A 2008 {\em Monte Carlo Methods\/} (Germany: Wiley-Vch
  Verlag GmbH \& Co.)

\end{thebibliography}


\providecommand{\newblock}{}

\end{document}